\documentclass[pra,twocolumn,superscriptaddress,aps,amsmath,amssymb,nofootinbib]{revtex4}
\usepackage{graphicx}
\usepackage{dcolumn}
\usepackage{bm}
\usepackage{color}

\usepackage[section]{placeins}
\usepackage{graphicx,epsfig}
\usepackage{amsmath}
\usepackage{amsfonts}
\usepackage{braket}
\usepackage{fancyhdr}
\usepackage{hyperref}
\usepackage[utf8]{inputenc}
\usepackage[T1]{fontenc}
\usepackage{amsthm}
\usepackage{hhline}
\def\beq{\begin{equation}}
\def\eeq{\end{equation}}
\def\beqa{\begin{eqnarray}}
\def\eeqa{\end{eqnarray}}

\begin{document}

\title{Quantum correlations and degeneracy of identical bosons in a 2D harmonic trap}
\date{\today}
\author{Pere Mujal}
\affiliation{Departament de F\'{i}sica Qu\`{a}ntica i Astrof\'{i}sica,
Universitat de Barcelona, Mart\'{i} i Franqu\`{e}s 1, 08028 Barcelona, Spain}
\affiliation{Institut de Ci\`{e}ncies del Cosmos (ICCUB), Universitat 
de Barcelona, Mart\'{i} i Franqu\`{e}s 1, 08028 Barcelona, Spain}
\author{Enric Sarlé}
\affiliation{Departament de F\'{i}sica Qu\`{a}ntica i Astrof\'{i}sica,
Universitat de Barcelona, Mart\'{i} i Franqu\`{e}s 1, 08028 Barcelona, Spain}
\author{Artur Polls}
\affiliation{Departament de F\'{i}sica Qu\`{a}ntica i Astrof\'{i}sica,
Universitat de Barcelona, Mart\'{i} i Franqu\`{e}s 1, 08028 Barcelona, Spain}
\affiliation{Institut de Ci\`{e}ncies del Cosmos (ICCUB), Universitat de 
Barcelona, Mart\'{i} i Franqu\`{e}s 1, 08028 Barcelona, Spain}
\author{Bruno Juli\'{a}-D\'{i}az}
\affiliation{Departament de F\'{i}sica Qu\`{a}ntica i Astrof\'{i}sica,
Universitat de Barcelona, Mart\'{i} i Franqu\`{e}s 1, 08028 Barcelona, Spain}
\affiliation{Institut de Ci\`{e}ncies del Cosmos (ICCUB), Universitat 
de Barcelona, Mart\'{i} i Franqu\`{e}s 1, 08028 Barcelona, Spain}
\affiliation{Institut de Ci\`{e}ncies Fot\`{o}niques, Parc Mediterrani 
de la Tecnologia, 08860 Barcelona, Spain}
\begin{abstract}
We consider a few number of identical bosons trapped in a 2D isotropic harmonic potential and also the $N$-boson system when it is feasible. The atom-atom interaction is modelled by means of a finite-range Gaussian interaction. The spectral properties of the system are scrutinized, in particular, we derive analytic expressions for the degeneracies and their breaking for the lower-energy states at small but finite interactions. We demonstrate that the degeneracy of the low-energy states is independent of the number of particles in the noninteracting limit and also for sufficiently weak interactions. In the strongly interacting regime, we show how the many-body wave function develops holes whenever two particles are at the same position in space to avoid the interaction, a mechanism reminiscent of the Tonks-Girardeau gas in 1D. The evolution of the system as the interaction is increased is studied by means of the density profiles, pair correlations and fragmentation of the ground state for $N=2$, $3$, and $4$ bosons.
\end{abstract}
\maketitle
\section{Introduction}

The problem of a particle trapped in a harmonic trap is one of the best known 
quantum systems. Going from a single particle to a system composed of $N$ 
interacting particles is, however, far more involved. Interestingly, recent 
advances in ultracold atomic gases have opened the possibility of studying 
systems of a few atoms, either fermions or bosons, trapped in potentials 
of different kind~\cite{Zurn,Bakr,Sherson,Islam}.

For the bosonic case, there are 
important results in 1D where the fermionization of the bosonic gas was preconized 
by Tonks and Girardeau~\cite{Girardeau} for the case of infinitely repulsive 
bosons and later confirmed experimentally in ultracold atomic gases~\cite{Paredes,Kinoshita}.
There are many works studying fermionization in 1D, for instance, in optical lattices \cite{Pupillo}, in few-atom mixtures \cite{Garcia-March4,Garcia-March,Pyzh}, for attractive interactions \cite{Tempfli} and for few dipolar bosons \cite{Deuretzbacher}. In other cases, the focus are quantum correlations \cite{Koscik,Barfknecht}, its effects in mixtures of distinguishable and identical particles \cite{Garcia-March3} and analytic ansatz to capture the physics in all interaction regimes \cite{Wilson}.

The case of two particles with contact interactions was considered 
in one, two and three dimensions in Ref.~\cite{Busch}. There, they obtained 
semi-analytic results finding the energies and wave functions as solution of 
transcendental equations. More general cases of few-body systems have 
been studied mostly in 3D, see Ref.~\cite{Blume_Report} and references therein. 

In 2D, semi-analytical approximate solutions to the case of two bosons with 
finite range interactions have been presented in Ref.~\cite{Doganov}. Other 
2D works include two and three-body exact solutions for fermions and bosons with contact interaction \cite{Liu}, fast-converging numerical methods for computing the energy spectrum for a few bosons \cite{Christensson}, the study of finite-range effects \cite{Imran,Imran2} and universality \cite{Kartavtsev,Shea}, condensation in trapped few-boson systems \cite{Daily}, and interacting few-fermions systems \cite{Stetcu,Liu2}.

In this paper, we study the properties for $N=2$, $3$, and $4$ identical bosons interacting through a finite-range interaction confined in a 2D 
isotropic harmonic trap by means of direct diagonalization of the Hamiltonian.

We analyze the properties of the system as we increase the strength of the 
interaction, going from the noninteracting regime to the 
strongly interacting one. In Sect.~\ref{secII}, we present the many-body Hamiltonian, 
including the two-body Gaussian-shaped interaction potential considered. We 
split the center-of-mass and relative parts of the Hamiltonian making use of 
Jacobi coordinates. In Sect.~\ref{secIII}, we consider first the noninteracting 
Hamiltonian and discuss in some details the degeneracies present in 
the many-body spectrum. In Sect.~\ref{secVI}, we focus on the effect of 
interactions on the many-body spectrum of the system. In Sect.~\ref{secVII}, we 
discuss the correlations which build in the ground state as the interaction 
is increased. Finally, the conclusions and summary are presented in Sect.~\ref{conclusions}.
\section{The $N$-boson Hamiltonian}
\label{secII}

We consider a system of $N$ identical bosons of mass $m$ trapped 
by an isotropic harmonic potential. The many-body Hamiltonian in 
first quantized form reads
\beq
{\cal H}=\sum_{i=1}^N 
\left ( -\frac{\hbar^2 }{2m}\nabla^2_i+\frac{1}{2}m\omega^2 \vec{x}^{\,2}_i\right )
+ g\, \sum_{i<j}^N V(\vec{x}_i-\vec{x}_j)\,.
\eeq
In usual ultracold atomic gases experiments, the atom-atom interactions 
are well approximated by a contact potential. In our case, we use a finite-size Gaussian potential,
\beq
\label{eqpot}
V(\vec{x}_i-\vec{x}_j)= g \; \frac{1}{\pi s^2}e^{-\frac{(\vec{x}_i-\vec{x}_j)^2}{s^2}}\,,
\eeq
where $g$ and $s$ characterize the strength and range of the interaction, respectively. Both parameters are considered to be tuneable. For instance, $g$ can be varied by means of a suitable Feshbach resonance. In the limit of $s$ going to zero, we recover a contact interaction with strength $g$. 
Regardless of $N$, we can split the Hamiltonian in two parts, 
${\cal H}={\cal H}_{\rm cm}+{\cal H}_{\rm r}$, using Jacobi coordinates, 
\beqa
\label{centerofmass}
\vec{R} &\equiv& \frac{1}{N}\sum_{i=1}^N \vec{x}_i \,, \\
\vec{r}_k&\equiv& \sqrt{\frac{2k}{k+1}} 
\left(\vec{x}_{k+1}-\frac{1}{k} \sum_{i=1}^k \vec{x}_i \right),\,\, k=1,\,...\,,N-1 \,. \nonumber
\eeqa
The center-of-mass part and relative part of the total Hamiltonian read
\beq
\label{eqcm}
{\cal H}_{\rm cm}=-\frac{\hbar^2}{2M}\nabla^2_{\vec{R}}+\frac{1}{2}M\omega^2\vec{R}^{\,2},
\eeq
\beq
\label{eqrel}
{\cal H}_{\rm r}=\sum_{k=1}^{N-1} \left(-\frac{\hbar^2}{2\mu}\nabla^2_{\vec{r}_k}
+\frac{1}{2}\mu \omega^2\vec{r}^{\,2}_k\right)+g\tilde{V}(\vec{r}_1,...\, ,\vec{r}_{N-1}),
\eeq
with the definitions $M\equiv Nm$ and $\mu\equiv m/2$. The interaction only 
appears in the relative part and takes the form
\beq
\begin{gathered}
\label{potentialN}
\tilde{V}(\vec{r}_1,...\, ,\vec{r}_{N-1})\equiv \\
 \sum_{i<j}^N V
\left(\vec{x}_i(\vec{R},\vec{r}_k,...\, ,\vec{r}_{N-1}) 
-\vec{x}_j(\vec{R},\vec{r}_k,...\, ,\vec{r}_{N-1})\right)\,.
\end{gathered}
\eeq
As a consequence, the change in the energy spectrum with increasing 
the interaction through $g$ or changing the range $s$ will come from 
a change in the energy associated to ${\cal H}_{\rm r}$.

\subsection{Second-quantized $N$-boson Hamiltonian}
\label{secIV}

Our numerical method to study the excitation spectrum will consist 
in truncating the Hilbert space of the $N$-boson system such that the 
particles can populate only the first $M$ single-particle eigenstates. We 
label the single particle states, $\psi_{i}(x,y)$, and their corresponding 
eigenenergies, $\epsilon_{i} = n_x+n_y+1$, with an index $i=1,\,...\,,M$ 
running through the pair of quantum numbers $n_x$ and $n_y$. With this 
truncation, the second quantized Hamiltonian reads
\beq
\label{eqMB1}
\hat{H}=\hat{H}_{0}+\hat{H}_{\rm int}\,.
\eeq
Where $\hat{H}_0$ and $\hat{H}_{\rm int}$ correspond to the single particle and 
interaction terms, 
\beqa
\label{eqMB2}
\hat{H}_{0}&=&\sum_{i=1}^M \hat{a}^{\dagger}_i \hat{a}_i \, \epsilon_i \,,\nonumber\\
\hat{H}_{\rm int}&=&\frac{g}{2} 
\sum_{i,j,k,l=1}^M \hat{a}^{\dagger}_i \hat{a}^{\dagger}_j \hat{a}_k \hat{a}_l \, V_{i,j,k,l} \,,
\eeqa
where 
\beqa
\label{eqMB4}
V_{i,j,k,l}&=&\frac{1}{\pi s^2}\int_{-\infty}^{\infty} dx \,dy\, dx'\, dy'  
\, \psi^{*}_{i}(x,y)\psi_{k}(x,y) 
\nonumber \\
&\times& 
\psi^{*}_{j}(x',y')\psi_{l}(x',y') \;e^{-\frac{(x-x')^2+(y-y')^2}{s^2}}.
\eeqa
The explicit analytical form of these integrals, $V_{i,j,k,l}$, is provided 
in Appendix~\ref{apintegrals}.

The operator $\hat{a}^{\dagger}_i$($\hat{a}_i$) creates(destroys) a particle in 
the single particle mode $i$, 
\beqa
\hat{a}^{\dagger}_i \ket{n_1,\, ... \, ,n_M}&=&\sqrt{n_i + 1}\ket{n_1,\, ... \, ,n_i+1,\, ... \, ,n_M}
\,,\nonumber\\
\hat{a}_i \ket{n_1,\, ... \, ,n_M}&=&\sqrt{n_i}\ket{n_1,\, ... \, ,n_i-1,\, ... \, ,n_M}\,.
\label{eqcrea}
\eeqa
They satisfy bosonic commutation relations,
$[ {\hat{a}}_i ,{\hat{a}}^{\dagger}_j ]=\delta_{i,j}$. 
We introduce the Fock basis,
\beq
\label{Fockbasis}
\ket{n_1,\, ... \, ,n_M}=\frac{(\hat{a}^{\dagger}_1)^{n_1} \dots 
(\hat{a}^{\dagger}_M)^{n_M}}{\sqrt{n_1 !\, ... \, n_M !}} \ket{\rm vac},
\eeq
where $\ket{\rm vac}\equiv \ket{0,\,...\,,0}$ is the vacuum state and, 
as we consider a fixed number of particles $N$, the quantum numbers 
$n_i$ verify
\beq
N=\sum_{i=1}^M n_i \,.
\label{eqNcons}
\eeq
The dimension of the Fock space is
\beq
D_N^M=\frac{(M+N-1)!}{(M-1)!N!},
\label{eqdimMB}
\eeq
which, for $N=2$, $3$, and $4$, gives, respectively, 
\beqa
D_2^M&=&\frac{M(M+1)}{2}  \,, \nonumber\\
D_3^M&=&\frac{M(M+1)(M+2)}{6} \,,  \nonumber\\
D_4^M&=&\frac{M(M+1)(M+2)(M+3)}{24} \,.
\eeqa

\section{Degeneracies in the noninteracting limit}
\label{secIII}

In this section, we will discuss the degeneracies present 
in the system in absence of interactions. First, we consider 
the two-boson case, in which the analysis is simpler, and then 
we shall explain the main degeneracies for $N$ bosons.

\subsection{The two-boson system}

In the noninteracting case, $g=0$, for the two-boson system, we 
can write down the Hamiltonian in second quantization, 
splitting the center of mass and the relative motion. Using from 
now on harmonic oscillator units, $\hbar \omega$ for energy and 
$ \sqrt{\hbar/ (m\omega)}$ for length, we have, in polar coordinates,
\beq
\label{eqboson1}
\hat{H}=\hat{H}_{\rm cm}+\hat{H}_{\rm r}= \hat{n}_{\rm cm} + \hat{n}_{r} + 2,
\eeq
where $\hat{H}_{\rm cm}=\hat{n}_{\rm cm} +1$, $\hat{H}_{r}=\hat{n}_{r} +1$. 
Therefore, we have a 2D harmonic oscillator for each part of the 
Hamiltonian. The corresponding eigenstates can be labelled as 
$\ket{n_{\rm cm},m_{\rm cm},n_{r},m_{r}}$, namely,
\beqa
\label{2deigenstates}
\hat{n}_{\rm cm}\ket{n_{\rm cm},m_{\rm cm},n_{r},m_{r}} &=&n_{\rm cm}\ket{n_{\rm cm},m_{\rm cm},n_{r},m_{r}},\nonumber\\
\hat{n}_{r}\ket{n_{\rm cm},m_{\rm cm},n_{r},m_{r}}&=&n_{r}\ket{n_{\rm cm},m_{\rm cm},n_{r},m_{r}},\nonumber\\
\hat{L}_{z,\rm cm}\ket{n_{\rm cm},m_{\rm cm},n_{r},m_{r}}&=&m_{\rm cm}\ket{n_{\rm cm},m_{\rm cm},n_{r},m_{r}},\nonumber\\
\hat{L}_{z,r}\ket{n_{\rm cm},m_{\rm cm},n_{r},m_{r}}&=&m_{r}\ket{n_{\rm cm},m_{\rm cm},n_{r},m_{r}},\nonumber\\
\eeqa
where $\hat{L}_{z,\rm cm}$ and $\hat{L}_{z,\rm r}$ are the third component 
of the center-of-mass orbital angular momentum and the relative orbital 
angular momentum, respectively, expressed in units of $\hbar$. However, 
those four quantum numbers have a restriction imposed by the symmetry 
of the wave function under the exchange of particles. The full wave 
function in polar coordinates for $\vec{R}$ and $\vec{r}$ reads
\beq
\label{eqboson2}
\begin{gathered}
\chi_{n_{\rm cm},m_{\rm cm},n_{r},m_{r}}\left(R,r,\varphi_R,\varphi_r\right)=
\nonumber \\
\chi_{n_{\rm cm},m_{\rm cm}}\left(\sqrt{2},R,\varphi_R\right) 
\chi_{n_{r},m_{r}}\left(\frac{1}{\sqrt{2}},r,\varphi_r\right),
\end{gathered}
\eeq
with 
\beq
\begin{split}
\label{eqpolarr}
\chi_{n,m}\left(\alpha, r,\varphi\right)=N_{n,m}\left(\alpha\right) e^{im\varphi}
\\
\times\, e^{-\frac{\left(\alpha r\right)^2}{2}} 
\left(\alpha r\right)^{|m|}L^{|m|}_{\frac{n-|m|}{2}}\left(\left(\alpha r\right)^2\right)\,.
\end{split}
\eeq
The $L^k_n(x)$ are the associated Laguerre polynomials defined as
\beq
\label{eqpolarr2}
L^k_n(x)\equiv\sum_{m=0}^n (-1)^m
\begin{pmatrix}
n+k \\
n-m
\end{pmatrix}
\frac{x^m}{m!}
\eeq
and $N_{n,m}\left(\alpha\right)$ is a normalization constant,
\beq
\label{eqpolarr3}
N_{n,m}\left(\alpha\right)=
\alpha\sqrt{\frac{\left(\frac{n-|m|}{2}\right)!}{\pi\left(\frac{n+|m|}{2}\right)!}}\,.
\eeq
\begin{table}[t]
\begin{centering}
\begin{tabular}{ |c|c|c|c|c|c|c|c| }
 \hline
 $n_{\rm cm}$  &  $n_r$  &  $m_{\rm cm}$  & $m_r$   &  $E$  &  $N_E$  & $d^b_{N_E}$  & $d^U_{N_E}$  \\[0.8ex]
 \hline
  0 &   0 &   0 &   0 &   2 &   0 &  1 &  0 \\
 \hline
  1 &   0 &  -1 &   0 &    &    &   &   \\
  1 &   0 &   1 &   0 &   3 &   1 & 2 &  0 \\
 \hline
  2 &   0 &  -2 &   0 &    &    &   &   \\
  2 &   0 &   0 &   0 &    &    &   &   \\
  2 &   0 &   2 &   0 &    &    &   &   \\
  0 &   2 &   0 &  -2 &   4 &   2 & 6  & 2   \\
  0 &   2 &   0 &   0 &    &    &   &   \\
  0 &   2 &   0 &   2 &    &    &  &   \\
 \hline
  3 &   0 &  -3 &   0 &    &    &   &   \\
  3 &   0 &  -1 &   0 &    &    &   &   \\
  3 &   0 &   1 &   0 &    &    &   &   \\
  3 &   0 &   3 &   0 &    &    &   &   \\
  1 &   2 &  -1 &  -2 &    &    &   &   \\
  1 &   2 &   1 &  -2 &   5 &  3& 10  & 4  \\
  1 &   2 &  -1 &   0 &    &    &   &   \\
  1 &   2 &   1 &   0 &    &    &   &   \\
  1 &   2 &  -1 &   2 &    &    &   &   \\
  1 &   2 &   1 &   2 &    &    &   &   \\
 \hline 
\end{tabular}
\end{centering}
\caption{Quantum numbers, energy, excitation energy number, 
degeneracy, and number of states with $m_r\neq0$ for the low-energy 
levels of a system of two noninteracting identical bosons 
trapped in a 2D isotropic harmonic potential. The energies 
are in units of $\hbar \omega$.}
\label{table1}
\end{table}
The wave function corresponding to the center of mass is symmetric 
under the exchange of particles, because $R$ and $\varphi_R$ remain 
unchanged upon exchanging particles 1 and 2, since 
$\vec{R}=\frac{1}{2}\left(\vec{x}_1+\vec{x}_2\right)$. However, the 
relative wave function is symmetric or antisymmetric depending on 
the quantum number $m_{r}$. We have defined the relative coordinate 
as $\vec{r}=\vec{x}_1-\vec{x}_2$, therefore the angle $\varphi_r$ 
changes to $\varphi_r + \pi$ and, due to the form of the wave function, see Eq.~(\ref{eqpolarr}), a factor $(-1)^{m_{r}}$ appears. For this 
reason, only the states with $m_{r}=$ even can describe the two-boson 
system. This implies that $n_{r}$ must also be an even number. To sum 
up, the four quantum numbers are
\beqa
\label{eqboson3}
\begin{cases}
n_{\rm cm}=0,1,2,3,4, \dots   \\
m_{\rm cm}=-n_{\rm cm},-n_{\rm cm}+2,\dots \, ,n_{\rm cm} \\
n_{r}=0,2,4,6,\dots \\
m_{r}=-n_{r},-n_{r}+2,\dots\, ,n_{r} \,.
\end{cases}
\eeqa
With the previous possible quantum numbers, we can determine the 
degeneracy for each energy level. We define the excitation energy 
number as the excitation energy per energy unit, $N_E\equiv E-E_0$. Then, the 
degeneracy for a given value of $N_E$ (see Appendix~\ref{apdeg}) is
\beqa
\label{eqboson4}
d^b_{N_E}&=&-\frac{1}{3}\left (\left \lfloor \frac{N_E}{2} \right \rfloor +1 \right)\\
&\times&
\left[4{\left \lfloor \frac{N_E}{2} \right \rfloor}^2 
+(2-3N_E)\left \lfloor \frac{N_E}{2} \right \rfloor -3 (N_E+1)\right] \,,\nonumber
\eeqa
where $\left \lfloor N_E/2 \right \rfloor$ indicates the floor function of $N_E/2$. The previous equation is valid for spinless bosons, which is the case considered in this work. However, for fermions and bosons with spin, the spatial antisymmetric states should be considered. The degeneracy for those states (see Appendix A) is
\beqa
\label{eqantisymmetric}
d^f_{N_E}&=&-\frac{1}{3}\left (\left \lfloor \frac{N_E}{2} \right \rfloor +1 \right)\\
&\times&
\left[4{\left \lfloor \frac{N_E}{2} \right \rfloor}^2 
+(8-3N_E)\left \lfloor \frac{N_E}{2} \right \rfloor -6 N_E\right] \,.\nonumber
\eeqa
Notice that the total degeneracy is given by \cite{Nathan},
\beq
d^{T}_{N_E}=d^b_{N_E}+d^f_{N_E}=\frac{(N_E+3)(N_E+2)(N_E+1)}{6}.
\eeq

\subsubsection{Unperturbed energy states}
\label{unperturbedstates}

We are also interested in knowing how many states have $m_{r}\neq0$ 
for each energy level, because these states are the ones that do 
not feel a zero-range interaction. For a finite but small range, 
these states are also expected to remain almost unperturbed for the considered range of interaction strengths. The number of states in each energy 
level such that their energy should not change significantly with a 
small Gaussian width (see Appendix~\ref{apdeg}) is
\beq
\label{eqboson5}
d^U_{N_E}=\left(-\frac{4}{3}\left \lfloor \frac{N_E}{2} \right \rfloor
+N_E+\frac{1}{3}\right)\left \lfloor \frac{N_E}{2} \right \rfloor
 \left(\left \lfloor \frac{N_E}{2} \right \rfloor+1 \right) \,.
\eeq

\subsection{$N$-boson system}
\label{nbs}

\begin{table}[t]
\begin{centering}
\begin{tabular}{|l|c|c|c|}
 \hline
 Eigenstates  &  $E$  &  $N_E$  & $d_{N_E}$  \\[.8ex]
 \hline
  $\ket{N,0,...\, ,0}$ & N &  0 &  1 \\
 \hline
  $\ket{N-1,1,0,...\, ,0}$ &  &  &   \\
  $\ket{N-1,0,1,0,...\, ,0}$ & N+1 & 1 &  2 \\
 \hline
  $\ket{N-1,0,0,1,0,...\, ,0}$ &  &  &   \\
  $\ket{N-1,0,0,0,1,0,...\, ,0}$ &  &  &   \\
  $\ket{N-1,0,0,0,0,1,0,...\, ,0}$ &  &  &   \\
  $\ket{N-2,2,0,...\, ,0}$ & N+2 & 2 & 6   \\
  $\ket{N-2,0,2,0,...\, ,0}$ &  &  &   \\
  $\ket{N-2,1,1,0,...\, ,0}$ &  &  &   \\
 \hline
  $\ket{N-1,0,0,0,0,0,1,0,...\, ,0}$ &  &   &   \\
  $\ket{N-1,0,0,0,0,0,0,1,0,...\,,0}$ &  &   &   \\
  $\ket{N-1,0,0,0,0,0,0,0,1,0...\,,0}$ &  &   &   \\
  $\ket{N-1,0,0,0,0,0,0,0,0,1,0,...\,,0}$ &  &   &   \\
  $\ket{N-2,1,0,1,0,...\, ,0}$ &  &   &   \\
  $\ket{N-2,0,1,1,0,...\, ,0}$ &  &   &   \\
  $\ket{N-2,1,0,0,1,0,...\, ,0}$ &  &   &   \\
  $\ket{N-2,0,1,0,1,0,...\, ,0}$ & N+3 & 3  & 14 \\
  $\ket{N-2,1,0,0,0,1,0,...\, ,0}$ &  &   &   \\
  $\ket{N-2,0,1,0,0,1,0,...\, ,0}$ &  &   &   \\
  $\ket{N-3,1,2,0,...\, ,0}$ &  &   &   \\
  $\ket{N-3,2,1,0,...\, ,0}$ &  &   &   \\
  $\ket{N-3,3,0,...\, ,0}$ &  &   &   \\
  $\ket{N-3,0,3,0,...\, ,0}$ &  &   &   \\
 \hline 
\end{tabular}
\end{centering}
\caption{Eigenstates expressed using the Fock basis (Eq.~(\ref{Fockbasis})), 
energy, excitation energy number and degeneracy, for the low-energy levels of 
a system of $N\geq 3$ noninteracting identical bosons trapped in a 2D isotropic 
harmonic potential. The energies are in units of $\hbar \omega$.}
\label{table2}
\end{table}

The procedure described above in order to compute the 
degeneracy is not valid for systems 
with more than two bosons. The reason is that we cannot label the 
symmetric (neither the antisymmetric) states under the exchange of 
a pair of particles using the previous quantum numbers. The 
symmetry of the relative Jacobi coordinates, defined in 
Eq.~(\ref{centerofmass}), under the exchange of two particles 
is not well defined. An alternative way for counting the degeneracy 
is by making use of the Fock basis introduced in the previous section, 
Eq.~(\ref{Fockbasis}). Those states are eigenstates of $\hat{H}_0$, i.e.,
\beqa
\label{energyfockbasis}
\hat{H}_0\ket{n_1,...\,,n_M}&=&
\left(\sum_{i=1}^M n_i\epsilon_i \right)\ket{n_1,...\,,n_M}\nonumber\\
&=&E\ket{n_1,...\,,n_M} \,.
\eeqa
The ground state of a system of $N$ identical spinless bosons 
in a 2D isotropic harmonic potential is always non-degenerate. In particular, for the noninteracting case, it corresponds to a state with all the bosons populating the non-degenerate single-particle ground state, i.e., the state $\ket{N,0,...\,,0}$. For 
any higher energy level of this system, labelled with $N_E=E-E_0$, there 
is a maximum number of degenerate states, $d^{\text{max}}_{N_E}$, 
that is reached when $N \geq N_E$.
\\
\\
{\bf Theorem.} $d_{N_E}=d^{\text{max}}_{N_E} \iff N \geq N_E$
\begin{proof}
From left to right, 
if we have reached $d^{\text{max}}_{N_E}$, one of the degenerate states is the one
with $N_E$ bosons in the single-particle states with excitation energy, 
$E^{\text{sp}}_{\text{exc}}=E^{\text{sp}}-E^{\text{sp}}_0=1$. Therefore, we have $N\geq N_E$ bosons. From right to left, if we have $N\geq N_E$ bosons, we have 
reached the maximum degeneracy because having less bosons would not allow us to 
have the previous discussed state, which is degenerate. Adding more 
bosons would not increase the number of degenerate states, since it is impossible to introduce new states with the same energy as the previous ones. This is due to the finite ways of decomposing $N_E$ as a sum of positive integers, without 
considering the order, that is, the number of partitions $p(N_E)$~\cite{Bruinier,Choliy}.
\end{proof}

Therefore, the degeneracy of the first $N_E+1$ energy levels is independent 
of the number of particles $N$ for any $N\geq N_E$. In Table~\ref{table2}, we give the low-energy states with their corresponding energies, excitation energy numbers and degeneracies for a system of $N$ bosons. In Table~\ref{table3}, we give $d^{\text{max}}_{N_E}$ for the first values of $N_E$. Computing the maximum degeneracy is analogous to computing the number of partitions of the integer $N_E$ where there are $n+1$ different kinds of part $n$ for $n = 1, 2, 3, ...$, \cite{oeis} and we can obtain it from its generating function,
\beq
\frac{1}{\prod_{k=1}^\infty(1-x^k)^{k+1}}=\sum_{N_E=0}^\infty d^{\text{max}}_{N_E} x^{N_E},
\eeq
and also,
\beq
d^{\text{max}}_{N_E}=\sum_{k=0}^{N_E} p(N_E-k)PL(k),
\eeq
where $PL(k)$ are the planar partitions of $k$ \cite{oeis2}.
Notice that the number of partitions is a lower bound of the maximum degeneracy,
\beq
d^{\text{max}}_{N_E} \geq p(N_E),
\eeq
and the equality would hold for non-degenerate single-particle states, e.g. for the 1D case.
\begin{table}[t]
\begin{centering}
\begin{tabular}{|c|c|c|c|}
 \hline
 \,\,\,\,\,$E$\,\,\,\,\, &  \,\,\,$N_E$\,\,\,  &  \,\,\,$p(N_E)$ \,\,\, & \,\,\, $d^{\text{max}}_{N_E}$ \,\,\,  \\ [0.8ex]
 \hline
    N      &  0   &  1  &   1   \\
 \hline
    N+1    &  1   &  1  &   2   \\
 \hline
    N+2    &  2   &  2  &   6   \\
 \hline
    N+3    &  3   &  3  &   14   \\
 \hline
    N+4    &  4   &  5  &   33   \\
 \hline
    N+5    &  5   &  7  &   70  \\
 \hline
    N+6    &  6   &  11  &  149    \\
 \hline
\end{tabular}
\end{centering}
\caption{Energy, excitation energy number, number of partitions of the 
excitation energy number and maximum degeneracy for the low-energy levels 
of a system of $N$ noninteracting identical bosons in a 2D isotropic 
harmonic potential. The maximum degeneracy, $d^{\text{max}}_{N_E}$, is equal 
to the degeneracy of the level $N_E$ if and only if $N \geq N_E$ 
(see the text for explanation).}
\label{table3}
\end{table}
%

\section{Energy spectra}
\label{secVI}

Our numerical method consists in the direct diagonalization of the 
truncated second-quantized Hamiltonian, as described in Sec.~\ref{secIV}. 
We will consider systems with $N=2$, $3$, and $4$ bosons. Direct diagonalization provides the energy spectrum of the Hamiltonian in the truncated space. In particular, we have used the {\it ARPACK} implementation of the Lanczos algorithm to obtain the lower part of the many-body spectrum.

\subsection{Two-boson energy spectrum}

\begin{figure}[t!]
\centering
\includegraphics[width=\columnwidth]{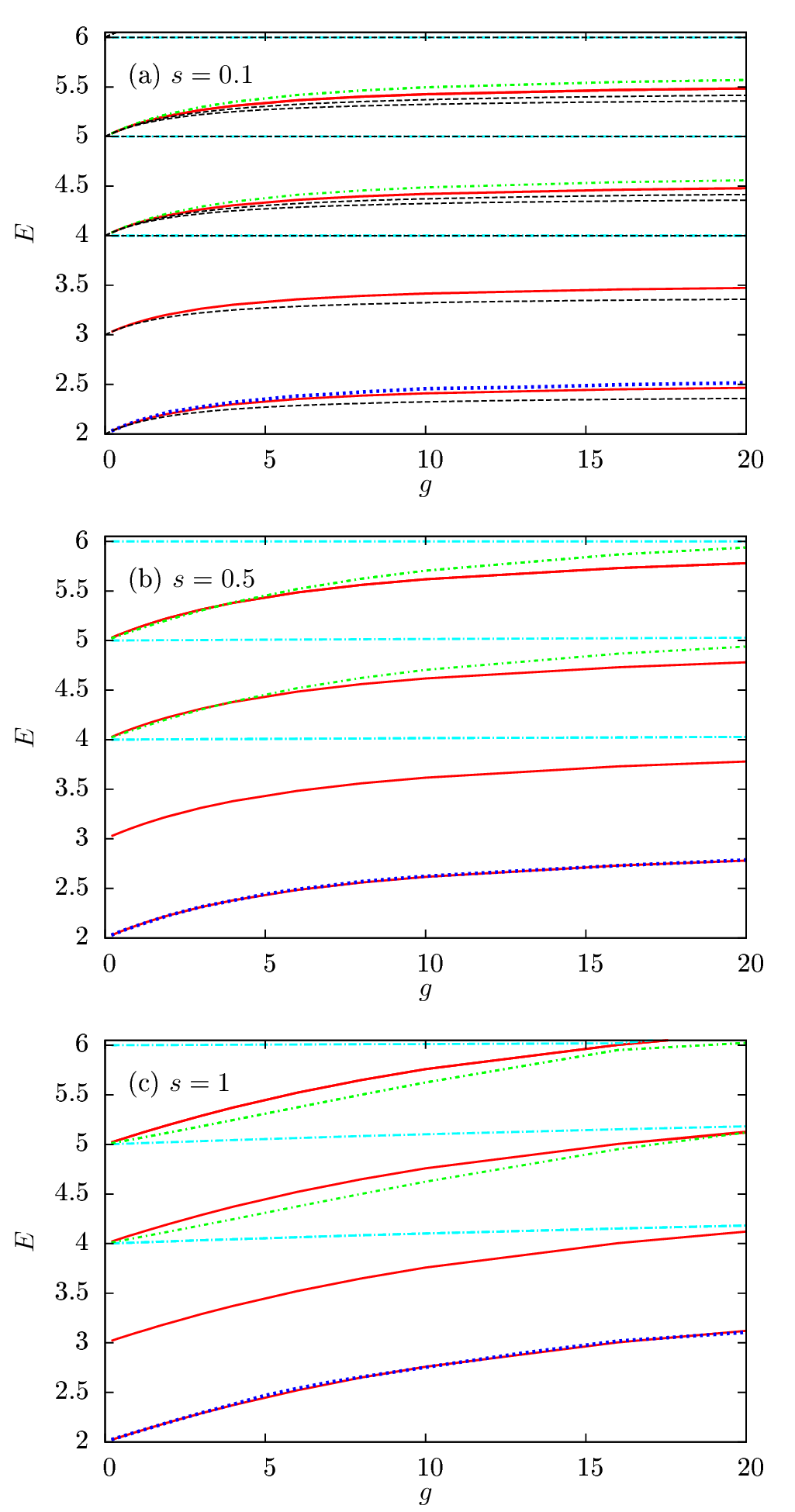}
\caption{(a-c) Low-energy spectrum for $N=2$ interacting bosons trapped in a 2D isotropic harmonic potential depending on the interaction strength $g$ for different values of the width $s$ of the two-body Gaussian-shaped potential. (Solid red lines) Energy of the ground state and the corresponding center-of-mass excitations. (Long-dashed pointed cyan lines) Unperturbed states. (Short-dashed pointed green lines) First relative excitation and the corresponding center-of-mass excitations. (a-c) (Blue dotted lines) Energy of the ground state computed with the variational ansatz of Eq.~(\ref{varmontecarlo2}). (Black dashed lines) Analytic approximate energy levels using Eq.~(17) of Ref.~\cite{Doganov} shown only in panel (a). Numerical results with (a) $M=200$ and $D_{2}^{200}=20100$, (b) and (c) $M=150$ and $D_{2}^{150}=11325$.}
\label{Fig:1}
\end{figure}

In Fig.~\ref{Fig:1}, we show the low-energy spectrum for the system of 
two interacting identical bosons in the harmonic trap. In the figure, 
we compare results obtained with three different values of 
$s=0.1$, $0.5$ and $1$. In all cases, the energy spectrum has a number 
of common features.

First, in the spectrum, there are the states discussed in Sect.~\ref{unperturbedstates}, which 
are essentially insensitive to the interaction. In the zero range limit, these 
are basically states with non-zero relative angular momentum, which do not 
feel the contact interaction~\cite{Doganov}. With finite interactions but for a
small range, $s=0.1$ and $0.5$, they remain mostly flat
for $g$ up to $20$. For $s=1$, their energy increases slightly with $g$, 
deviating from the zero range prediction.

Second, the ground state energy 
increases linearly with $g$ for small values of $g$. Up to first order perturbation theory, the energy is given by
\beq
\label{pertgroundN2}
E_0\simeq 2 + \frac{g}{\pi\left(s^2+2\right)}.
\eeq
However, the ground state energy tends to 
saturate as $g$ is increased. For smaller values of $s$, this saturation 
takes place at smaller values of $g$.

Third, there are the energies coming from the relative part of the Hamiltonian with the center of mass at the ground state, i.e. $n_{\rm cm}=0$. The ground state is one of these states and there is one state of this type in each energy level with an even $N_E$ in the noninteracting limit.

Finally, the spectrum also contains center-of-mass excitations~\cite{Busch}, which are easily 
recognized as constant energy shifts independent of $g$ with respect to states with $n_{\rm cm}=0$.

For comparison, we depict also the approximate values of~\cite{Doganov} in panel (a) of Fig.~\ref{Fig:1}. As 
reported in Ref.~\cite{Doganov}, their approximate solution -- which is 
not variational -- starts to deviate from the exact numerical results at 
values of $g\simeq 4$. The approximation gives, however, a fairly good overall 
picture of the low-lying two-particle spectrum. 

\subsection{Degeneracy for the interacting two-boson system}

 We can label the states with three quantum numbers. Two are the ones corresponding to the center of mass, $n_{\rm cm}$ and $m_{\rm cm}$, and the other is a new quantum number, $\nu_r$, that labels the nondegenerate eigenstates of the relative part of the Hamiltonian. We can write those states as
\beq
\label{statesinteracting}
\Psi(R,\varphi_R,r)=\chi_{n_{\rm cm},m_{\rm cm}}(\sqrt{2},R,\varphi_R)f_{\nu_r}(r),
\eeq
where $\chi_{n_{\rm cm},m_{\rm cm}}(\sqrt{2},R,\varphi_R)$ is given in Sect.~\ref{secIII} and $f_{\nu_r}(r)$ is the relative wave function, that depends on $g$ and $s$. The other states that are in the spectrum are the unperturbed ones (almost unaffected by the interaction). Their degeneracy is given in Sec.~\ref{secIII}. The states of Eq.~(\ref{statesinteracting}), for a given $\nu_r$, are degenerate with degeneracy given by the 2D harmonic oscillator of the center-of-mass part, i.e., their degeneracy is $n_{\rm cm}+1$. From each noninteracting energy level with even $N_E$, a state with a new $\nu_r$ arises, and its center-of-mass excitations appear in higher energy levels with degeneracy $n_{\rm cm}+1$, too.
To sum up, the ground state is nondegenerate. The first excited state is two-degenerate and the two states are the two possible center-of-mass excitations of the ground state. The third noninteracting energy manifold (6 states with $E(g=0)=4$) splits in three: 1) three center-of-mass excitations of the ground state, 2) two unperturbed states and, 3) the new relative state with quantum numbers $n_{\rm cm}=0$, $m_{\rm cm}=0$ and $\nu_r=1$ with $E(g=2)=4.21$.
We give the degeneracy and the quantum numbers of the low-energy states in Table~\ref{table4}.
\begin{table}[t]
\begin{centering}
\begin{tabular}{ |c|c|c|c|c|c|c|c| }
 \hline
 $n_{\rm cm}$  &  $n_r$  &  $m_{\rm cm}$  & $m_r$   &  $\nu_r$ & $E(g=0)$  &  $E(g=2)$  & $d_{int}(g=2)$  \\[0.8ex]
 \hline
  0 &   - &   0 &   0 &   1 &   2 &  2.23 &  1 \\
 \hhline{|=|=|=|=|=|=|=|=|}
  1 &   - &  -1 &   0 &   1 &   3 & 3.23  &   \\
  1 &   - &   1 &   0 &   1 &   3 & 3.23 &  2 \\
 \hhline{|=|=|=|=|=|=|=|=|}
  2 &   - &  -2 &   0 &   1 &   4 & 4.23  &   \\
  2 &   - &   0 &   0 &   1 &   4 & 4.23  &  3 \\
  2 &   - &   2 &   0 &   1 &   4 & 4.23  &   \\
  \hline
  0 &   - &   0 &   0 &   2 &   4 & 4.21  &  1 \\
  \hline
  0 &   2 &   0 &  -2 &   - &   4 & 4.00  &    \\
  0 &   2 &   0 &   2 &   - &   4 & 4.00 &  2 \\
 \hhline{|=|=|=|=|=|=|=|=|}
  3 &   - &  -3 &   0 &   1 &   5 & 5.23  &   \\
  3 &   - &  -1 &   0 &   1 &   5 & 5.23  &   \\
  3 &   - &   1 &   0 &   1 &   5 & 5.23  & 4  \\
  3 &   - &   3 &   0 &   1 &   5 & 5.23  &   \\
\hline  
  1 &   - &  -1 &   0 &   2 &   5 & 5.21  &   \\
  1 &   - &   1 &   0 &   2 &   5 & 5.21  & 2  \\
\hline  
  1 &   2 &  -1 &  -2 &   - &   5 & 5.00  &   \\
  1 &   2 &   1 &  -2 &   - &   5 & 5.00  &   \\
  1 &   2 &  -1 &   2 &   - &   5 & 5.00  & 4  \\
  1 &   2 &   1 &   2 &   - &   5 & 5.00  &   \\
 \hline 
\end{tabular}
\end{centering}
\caption{Quantum numbers, energy in the noninteracting limit, energy at $g=2$ and 
degeneracy, for the low-energy 
levels of a system of two interacting identical bosons 
trapped in a 2D isotropic harmonic potential. The energies 
are in units of $\hbar \omega$ and the ones with $g=2$ correspond to a vertical cut in Fig.~\ref{Fig:1} panel (b), $s=0.5$.}
\label{table4}
\end{table}
\subsection{Three and four-boson energy spectra}
\begin{figure}[t]
\centering
\includegraphics[width=\columnwidth]{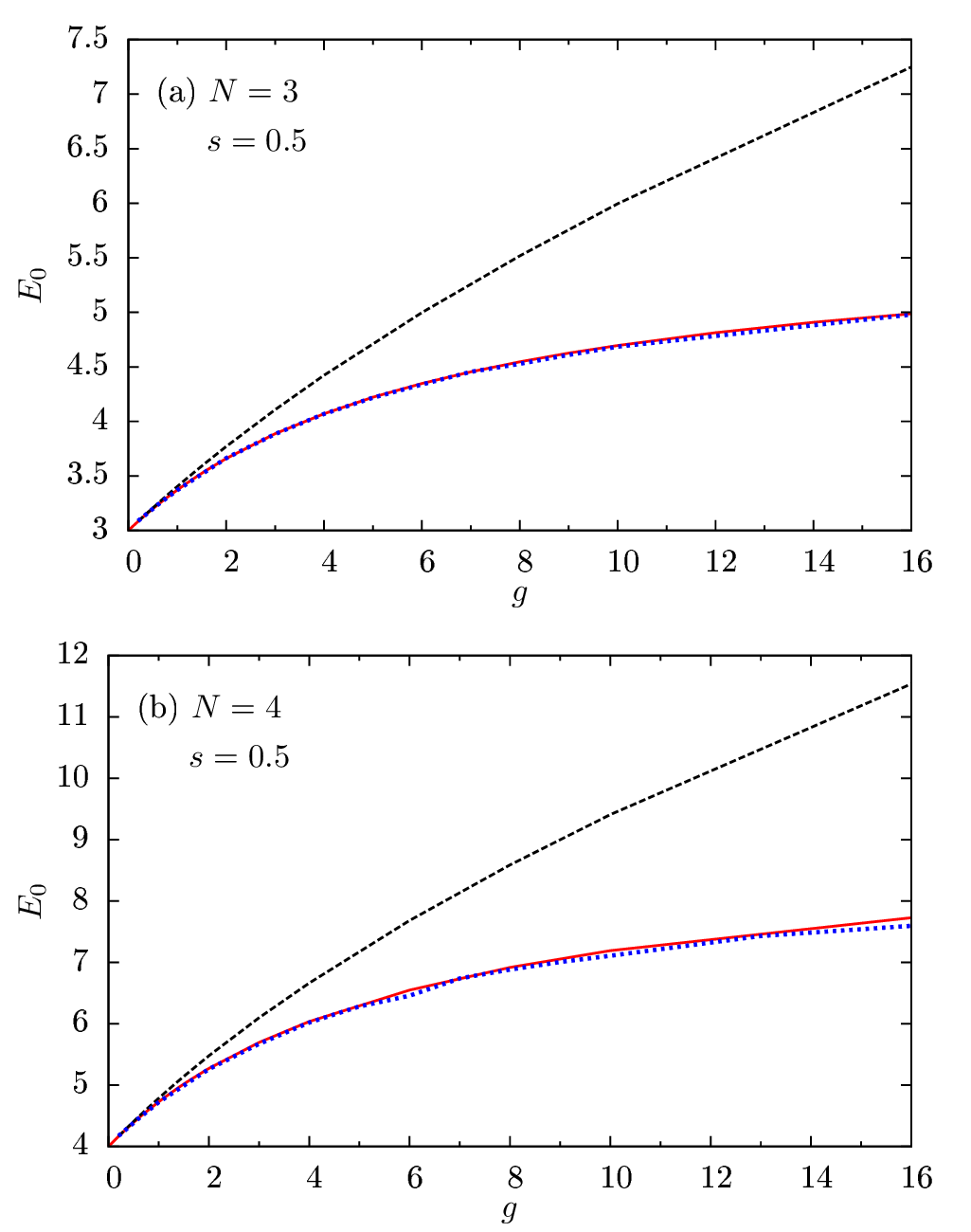}
\caption{Ground-state energy for (a) N=3 and (b) N=4 interacting bosons 
trapped in a 2D isotropic harmonic potential depending on the interaction 
strength $g$. (Red solid line) computed numerically with {\it ARPACK} 
and (a) $M=80$ and (b) $M=50$, (blue dotted line) computed with the variational many-body wave function of Eq.~(\ref{varmontecarlo2}), (black dashed line) computed with a Gaussian
variational ansatz, Eq.~(\ref{varansatz}).}
\label{Fig:2}
\end{figure}
\begin{figure}[t]
\centering
\includegraphics[width=\columnwidth]{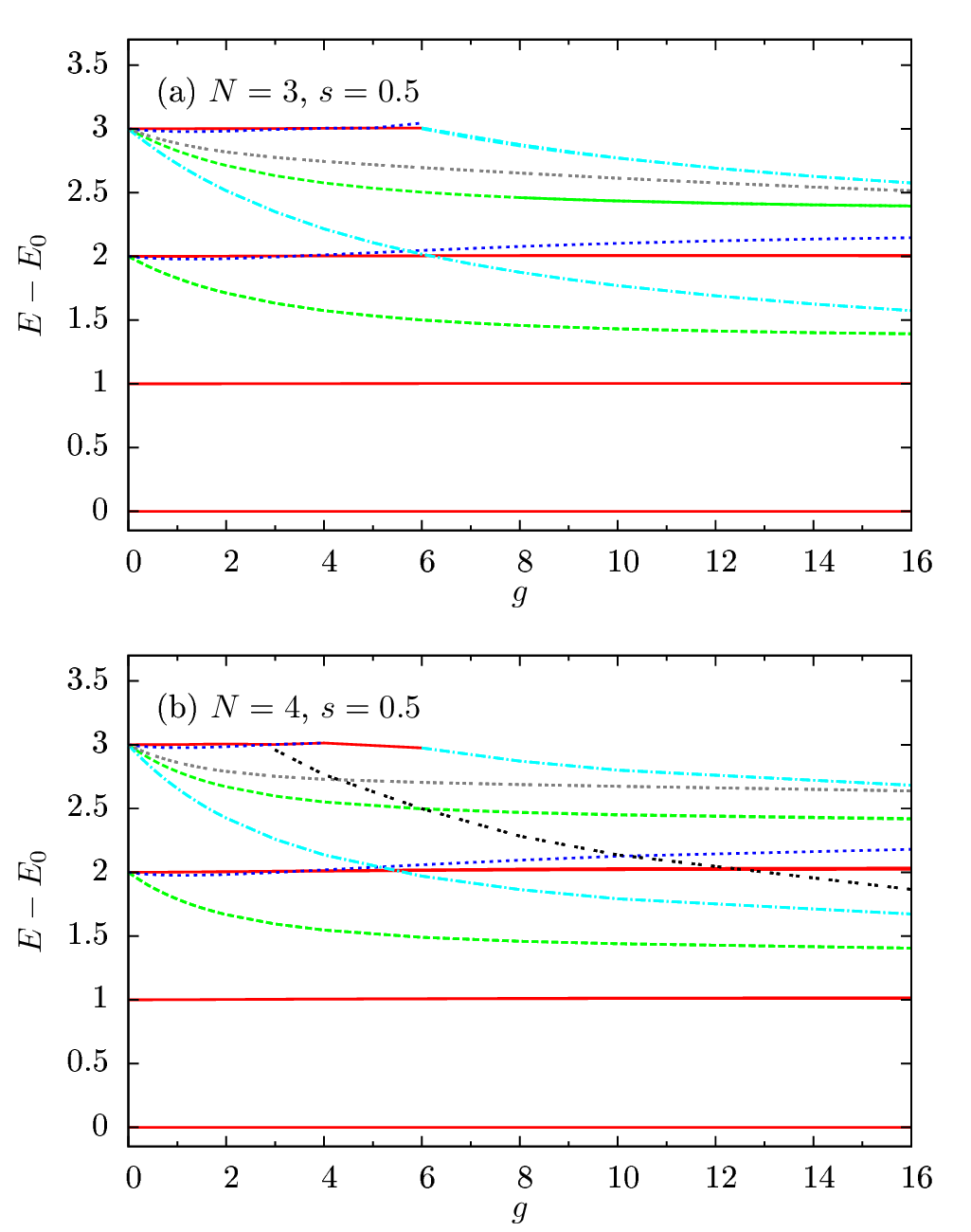}
\caption{Low-energy spectrum for (a) $N=3$ and (b) $N=4$ interacting bosons 
trapped in a 2D isotropic harmonic potential depending on the interaction 
strength $g$. (Red solid lines) Energy of the ground state, (green dashed lines) the first, (blue dotted lines) the second, (cyan dashed-dotted lines) the third, (grey triple-dotted lines) the fourth and (black double-dotted lines) the fifth relative excitations and, respectively, their center-of-mass excitations in the same kind of line and color. The number of modes that we have used is (a) $M=80$ and (b) $M=50$, that corresponds to Hilbert spaces of dimension $D_3^{80}=88560$ and $D_4^{50}=292825$, respectively.}
\label{Fig:3}
\end{figure}

Our exact diagonalization scheme allows us to obtain the lowest 
part of the many-body spectrum for systems of up to $4$ bosons with 
good accuracy, up to values of $g\simeq 20$. In Fig.~\ref{Fig:2} 
we report the ground state energy for $N=3$ and $N=4$ bosons compared 
with a simple mean-field variational ansatz using the following 
wave function,
\beq
\label{varansatz}
\Psi(\vec{x}_1,\,...\,,\vec{x}_N)
=\left(\frac{\alpha}{\pi}\right)^{\frac{N}{2}} \prod_{i=1}^N  e^{-\frac{1}{2}\alpha \vec{x}^2_i},
\eeq
and finding the optimum $\alpha^{*}$ that minimizes the energy
\beqa
E_0(\alpha)&=&\int_{-\infty}^{\infty}d\vec{x}_1 \, ... \, d\vec{x}_N \,
\Psi^{*}(\vec{x}_1,\,...\,,\vec{x}_N)H\Psi(\vec{x}_1,\,...\,,\vec{x}_N)
\nonumber \\
&=&N\left(\frac{\alpha}{2}+\frac{1}{2\alpha}\right)+\frac{gN(N-1)\alpha}{2\pi\left(\alpha s^2+2\right)}.
\eeqa

As expected, this mean-field ansatz captures well the behaviour of the 
ground state of the system for small values of $g$. For $g\simeq 2$, 
however, we already observe substantial deviations, with the meanfield 
prediction overstimating the ground state energy considerably. In 
particular, as we will see below, the system develops strong 
beyond-mean-field correlations as $g$ is increased.

In addition, we introduce a two-body-correlated variational many-body ansatz of Jastrow type~\cite{Jastrow},
\beq
\label{varmontecarlo2}
\Psi(\vec{x}_1,\,...\,,\vec{x}_N)=\left(\frac{\alpha}{\pi}\right)^{\frac{N}{2}} 
\prod_{i=1}^N  e^{-\frac{1}{2}\alpha \vec{x}^2_i}\prod_{j<i}^N\left(1-a e^{-b(\vec{x}_i -\vec{x}j)^2}\right),
\eeq
where $\alpha$, $a$ and $b$ are the variational parameters. We observe in Fig.~\ref{Fig:2} that the energies computed with this ansatz, using standard Monte-Carlo methods, are very close, and some times even below, the exact diagonalization ones. To improve the latter, one needs to enlarge the Hilbert space (larger $M$) to get a slightly lower upper bound. In principle, the exact diagonalization procedure for a given $M$ provides an upper bound for the ground state and each excited state. In the next section, we explain the physical interpretation of the variational parameters and discuss how well the ansatz captures the physics of the problem.

The low-energy spectrum for $N=3$ and $N=4$ at smaller values of $g$ is fairly similar. This is not unexpected as the degenerate manifolds are the same irrespective of the number of particles, see Sect.~\ref{nbs}. The first excited state is a center-of-mass excitation, the Kohn mode, as seen clearly in the excitation spectra shown in Fig.~\ref{Fig:3}. 

Even for $g$ up to $16$, the low-energy spectra for $N=3$ and $N=4$ are quite similar. The overall picture is qualitatively the same for both cases, although for $N=4$ there are extra levels crossing. In Fig.~\ref{Fig:3} panel (b), there is a level that starts crossing the highest energy level depicted at $g\simeq 3$. This line in the spectrum comes from the fourth excited level in the noninteracting limit and is also expected to appear for systems with more particles, e.g. $N=5$. It arises from the existence of a degenerate kind of states that are found only for $N\geq4$, as it is explained in Sect.~\ref{secIII}.

\subsection{Degeneracy for the interacting three and four-boson systems}

One major difference for more than two particles, is that we do not find states not affected by the interaction. Moreover, the degeneracy of the eigenfunctions of the relative part of the Hamiltonian is not $1$. Therefore, the states cannot be uniquely characterized by $\nu_r$. However, we can identify the states that are center-of-mass excitations of lower energy states. In Fig.~\ref{Fig:3}, in both panels, for example, for $g=1$, we know the degeneracy of all the energy levels and we can identify them. The ground state is nondegenerate. As we have said before, the first excited state is a center-of-mass excitation, with degeneracy 2. The second excited state decomposes in three states corresponding to the next center-of-mass excitations of the ground state, there are two degenerate states corresponding to a relative excitation, and finally a different relative excitation. The third excited energy level in the noninteracting limit splits when $g$ is increased in the next center-of-mass excitations of the states of the previous level, i.e., four center-of-mass excitations of the ground state, four center-of-mass excitations of the previous two-degenerate relative excited states, and two more degenerate states corresponding to two center-of-mass excitations of the single-degenerate relative energy level that appeared in the second excited state when $g$ was increased. Moreover, there are two pairs of different relative excited states that split from the noninteracting third energy level. This behaviour is the same independently of $N$ for $g$ sufficiently small, for instance, for $N=4$ up to $g=3$, where there is the previous discussed crossing of levels.

\subsection{$N$-boson energies up to first order in perturbation theory}
Using the analytic expressions of the integrals of the interaction that are given in Appendix \ref{apintegrals}, we compute the energies of the first three energy levels in first order perturbation theory. For the ground state of the system, the energy is given by
\beq
\label{groundpert}
E_0\simeq N+g\frac{N(N-1)}{2\pi\left(s^2+2\right)}.
\eeq
The next level has energy
\beq
\label{1pert}
E_1\simeq N+1+g\frac{N(N-1)}{2\pi\left(s^2+2\right)}.
\eeq
The third energy level splits in three, in the way that is discussed in the previous section that is also valid for $N$ bosons, with energies
\beq
\label{2pert1}
E_{2_1}\simeq N+2+g\frac{N(N-1)}{2\pi\left(s^2+2\right)},
\eeq
\beq
\label{2pert2}
E_{2_2}\simeq N+2+g\frac{N\left(N(2+s^2)^2-s^2(8+s^2)-8\right)}{2\pi\left(s^2+2\right)^3},
\eeq
\beq
\label{2pert3}
E_{2_3}\simeq N+2+g\frac{N\left(N(2+s^2)^2-s^2(8+s^2)-4\right)}{2\pi\left(s^2+2\right)^3}.
\eeq

The similarity in the energy difference, $E-E_0$, for the case of $N=3$ and $N=4$ plotted in Fig.~\ref{Fig:3} for a small $g$ can be understood using the previous expressions. The corresponding excitation energies are, in this approximation,
\beq
\label{dif1}
E_1-E_0=1,
\eeq
\beq
\label{dif21}
E_{2_1}-E_0=2,
\eeq
\beq
\label{dif22}
E_{2_2}-E_0=2-g\frac{2N\left(1+s^2\right)}{\pi\left(s^2+2\right)^3},
\eeq
\beq
\label{dif23}
E_{2_3}-E_0=2-g\frac{2Ns^2}{\pi\left(s^2+2\right)^3}.
\eeq

In the first two cases, Eq.~(\ref{dif1}) and Eq.~(\ref{dif21}), we recover the first and the second center-of-mass excitations that are red solid lines in Fig.~\ref{Fig:3}. The presence of the factor $N$ in the quantity $E_{2_2}-E_0$, see Eq.~(\ref{dif22}), explains why the slope of the green dashed lines is slightly bigger in absolute value for $N=4$, panel (b), than for $N=3$, panel (a), in Fig.~\ref{Fig:3} for $g\simeq0$. This effect would be notorious when comparing the spectrum for two very different numbers of particles. Finally, we also see that the second term in $E_{2_3}-E_0$ is proportional to $N$, but in that case, for $s$ small, the second term becomes negligible. Therefore, the blue dotted lines are very close to the red solid lines in the spectra for $g\simeq0$, as we have used $s=0.5$. In the zero-range limit, this approximation gives $E_{2_3}(s \rightarrow 0)=E_{2_1}(s \rightarrow 0)$.
As the perturbative correction affects only the relative motion, the corrections to $E_0$, $E_1$ and $E_{2_1}$ are equal. 

\section{Interactions and quantum correlations}
\label{secVII}

\begin{figure}[t]
\centering
\includegraphics[width=\columnwidth]{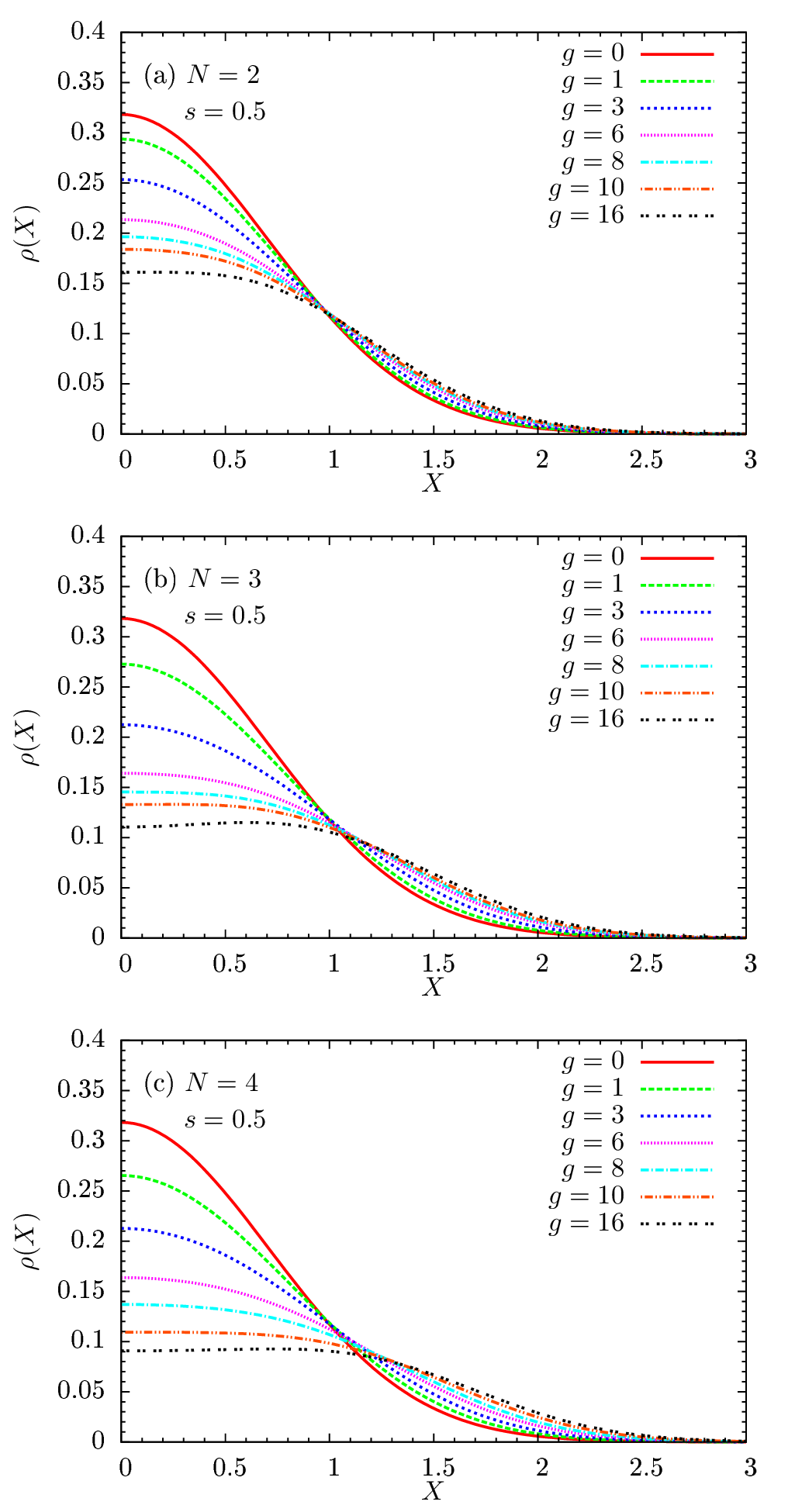}
\caption{Density profile of the ground state for (a) $N=2$, (b) $N=3$ and (c) 
$N=4$ interacting bosons trapped in a 2D isotropic harmonic potential for 
different values of the interaction strength $g$ for a fixed range $s=0.5$. 
The number of modes that we have used is $M=50$, which corresponds to a Hilbert space with dimension (a) $D_2^{50}=1275$, (b) $D_3^{50}=22100$ and (c) $D_4^{50}=292825$.}
\label{Fig:4}
\end{figure}

\begin{figure}[t]
\centering
\includegraphics[width=\columnwidth]{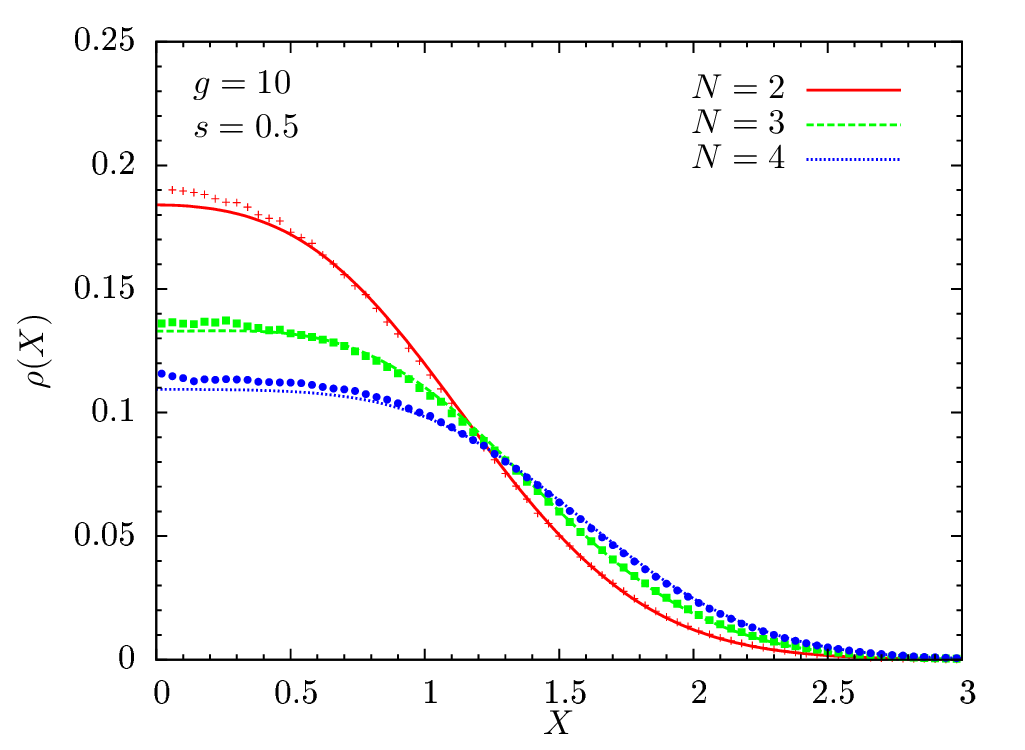}
\caption{Density profile for $N=2$, $3$, and $4$ interacting bosons trapped 
in a 2D isotropic harmonic potential computed with the ground state obtained 
using {\it ARPACK} with $M=50$ (lines) and obtained from the variational ansatz, Eq.~(\ref{varmontecarlo2}),
(crosses, squares and dots) for $g=10$ and $s=0.5$.}
\label{Fig:5}
\end{figure}

%

As seen in the previous section, the ground state energy of the system for 
$N=2, 3$ and 4 tends to saturate as we increase the strength of the atom-atom
interactions. This saturation starts to occur for values $g$ for which the 
mean-field variational ansatz starts to deviate from the exact results. This 
reminds of a similar effect found in 1D systems, where the ground state evolves 
from mean-field to Tonks-Girardeau gas as the interaction strength is increased. 
In the Tonks-Girardeau limit, the atoms do avoid completely the atom-atom contact 
interaction by building strong correlations which in 1D are easily understood 
from the Bose-Fermi mapping theorem \cite{Girardeau2}. In 2D, no such mapping exist. However, we expect that the system should build suitable correlations to avoid the 
interaction, resulting in a saturation of the energy for increasing $g$. 

For the ground state, besides the exact diagonalization method, we have also made use of a correlated 
variational ansatz, Eq.~(\ref{varmontecarlo2}), to enlighten the discussion. The energies and properties associated to this variational ansatz are evaluated by means of Monte-Carlo methods (standard Metropolis algorithm).
The physical meaning of the variational parameters
is quite transparent. $\alpha$ directly affects the overall size of the cloud. 
The two-body Jastrow correlations are parameterized by $a$ and $b$. 
Two limiting cases are illustrative. If the system is fully condensed we 
will have $a=0$, while $a=1$ would correspond to building a zero of the wave 
function whenever two atoms are at the same position. $b$ affects the two-body 
correlation length. Thus, we expect the following behavior: for values of $g\simeq 0$ 
we should have $a=0$ ($b$ is thus irrelevant) and $\alpha$ close to $1$. For increasing $g$, $\alpha$ decreases to avoid the interaction by simply putting the atoms apart. As we increase $g$, two-body correlations build in, $a\neq 0$ and $\alpha$ should stop 
decreasing as the correlation is more efficient to separate the atoms.

Let us first discuss the density profile of the clouds, see Appendix~\ref{appendixdenspair} for definitions. In Fig.~\ref{Fig:4} 
we show the density profile, normalized to unity, depending on the radial coordinate 
$X=\sqrt{x^2+y^2}$, computed with our exact diagonalization procedure. 
Due to the symmetry of the trap, the density profile of the ground state does not have angular 
dependence, see Appendix~\ref{appendixdenspair}, Eq.~(\ref{densityyy}). 
In panels (a), (b) and (c) we show results for $N=2$, $3$, and $4$. 
In all cases, with the same value of $s=0.5$. We compare densities 
obtained for different values of $g$.

Irrespective of $N$ we observe a 
number of common features. For $g=0$, the system has a Gaussian density 
profile which, as $g$ is increased, evolves into a profile with a flat 
region for $X\leq 1$ at $g\simeq 16$. As $N$ is increased, the size of 
the inner plateau increases, thus tending towards an homogeneous density.

The quality of our variational approach is seen in Fig.~\ref{Fig:5}. 
We compare density profiles obtained with the exact diagonalization procedure 
with those obtained variationally by means of Eq.~(\ref{varmontecarlo2}). 
As seen in the figure, the variational wave function provides a fairly 
accurate representation of the density profile for $N=2$, $3$, and $4$. In particular, 
it captures well the appearance of the plateau. 
%

\begin{figure}[t]
\centering
\includegraphics[width=\columnwidth]{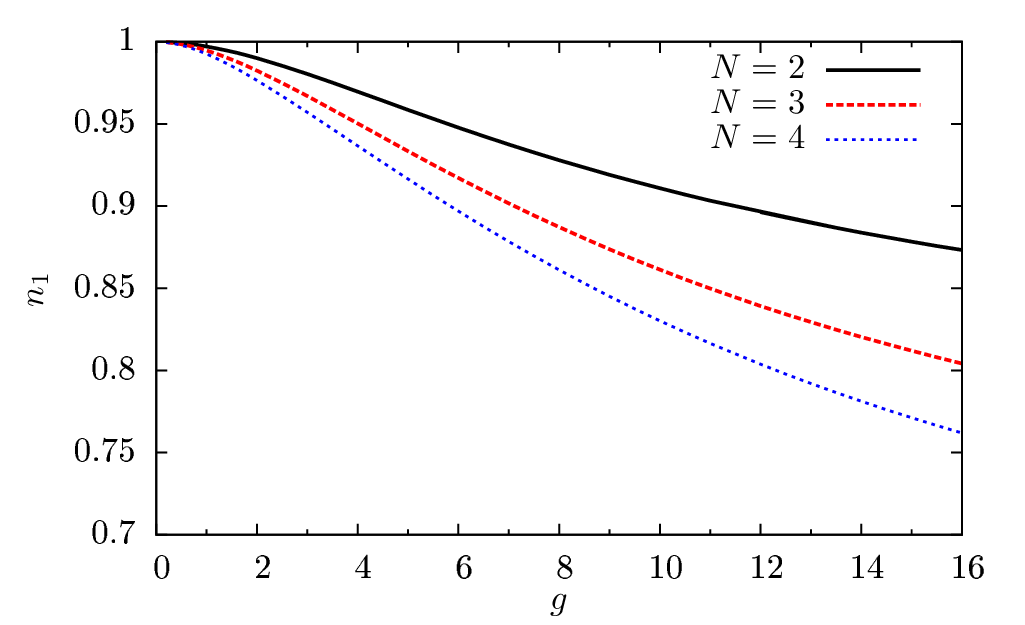}
\caption{Condensed fractions of the ground state for (black line) $N=2$, 
(red-dashed line) $N=3$ and (green spotted line) $N=4$ interacting bosons 
trapped in a 2D isotropic harmonic potential depending on the interaction 
strength $g$ for a fixed range $s=0.5$. The number of modes that we 
have used is $M=50$ and the rest of the eigenstates of the one-body density matrix are much more smaller than the biggest one.}
\label{Fig:6}
\end{figure}

The effect of increasing the interaction among the atoms is manifold. 
As we have seen above, the density profile is modified and the gas 
becomes close to homogeneous in the inner part of the trap. This change 
in the density is however accompanied by a change in the correlations 
present in the system. Actually, the gas goes from a fully condensed 
state to a largely fragmented one as we increase the interaction. In 
Fig.~(\ref{Fig:6}), we depict how the condensed fraction for $N=2$, $3$, and $4$ decreases when increasing the interaction 
strength. For the same value of $g$, the fragmentation in the system is 
larger for larger number of particles.

\begin{figure}[t!]
\centering
\includegraphics[width=\columnwidth]{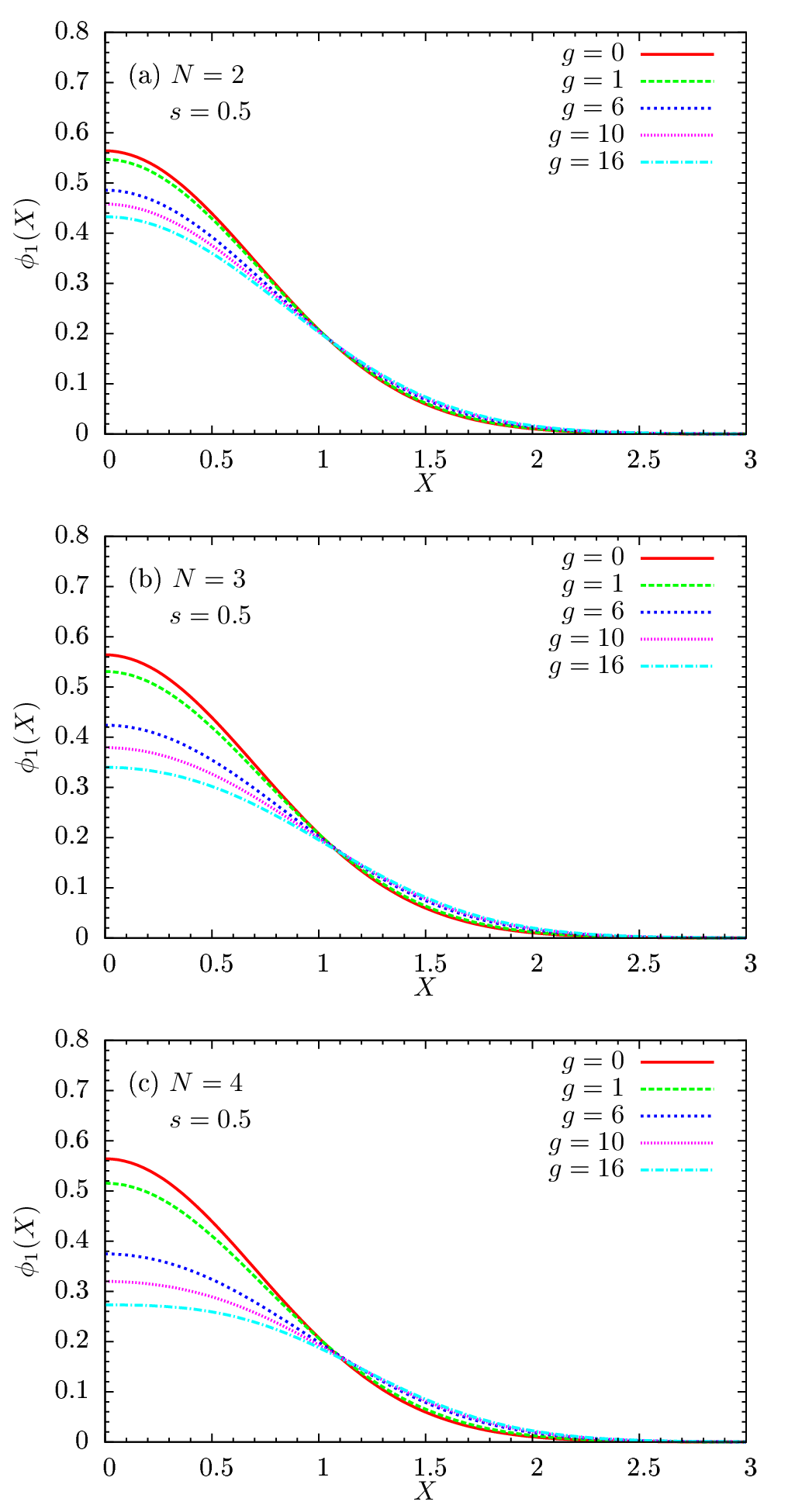}
\caption{Single-particle eigenstate of the one-body density matrix in which the particles condense. We use Eq.~(\ref{eigenwavedens}) and the values of $C_0$ and $C_1$ computed numerically diagonalizing the one-body density matrix, Eq.~(\ref{eqonebody}), for different values of $g$. (a) $N=2$ bosons, (b) $N=3$ and (c) $N=4$. The fraction of condensed particles is plotted in Fig.~\ref{Fig:6}. The number of modes that we have used is $M=50$, which corresponds to a Hilbert space with dimension (a) $D_2^{50}=1275$, (b) $D_3^{50}=22100$ and (c) $D_4^{50}=292825$.}
\label{Fig:7}
\end{figure}

\begin{figure}[t!]
\centering
\includegraphics[width=\columnwidth]{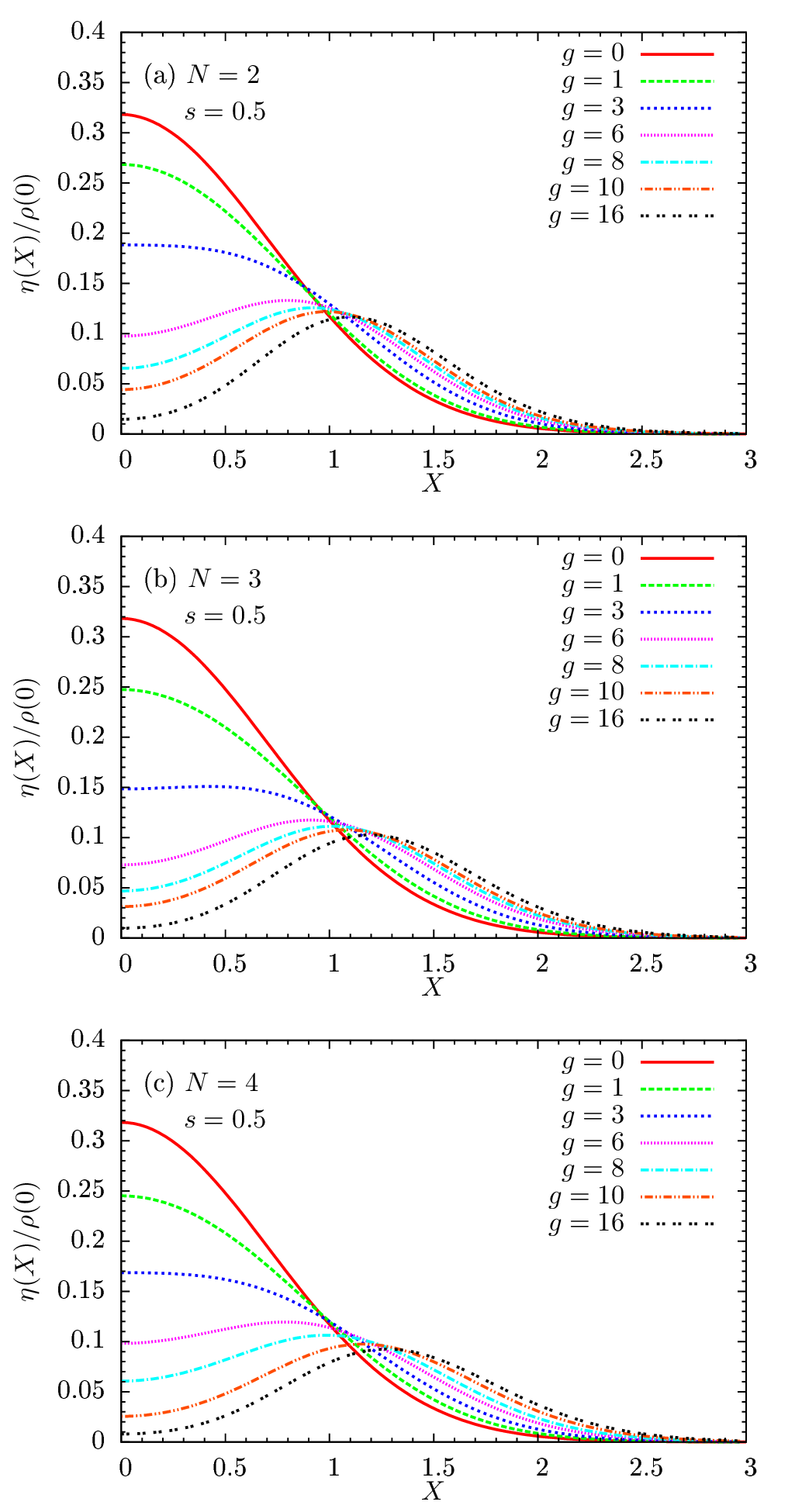}
\caption{Probability density, $\eta(X)/\rho(0)$, of finding a particle 
at position $X$ once we have found one particle at the origin, $X=0$, for 
(a) $N=2$, (b) $N=3$ and (c) $N=4$ interacting bosons trapped in a 2D 
isotropic harmonic potential for different values of the interaction 
strength $g$ for a fixed range $s=0.5$. The number of modes that we have used is $M=50$, which corresponds to a Hilbert space with dimension (a) $D_2^{50}=1275$, (b) $D_3^{50}=22100$ and (c) $D_4^{50}=292825$.}
\label{Fig:8}
\end{figure}

The most populated eigenstate of the one-body density matrix (natural orbit), is found to have the approximate form, using the $\ket{n_x,n_y}$ basis,
\beq
\label{eigenvectordens}
\ket{\phi_1}\simeq C_0\ket{0,0}+C_1\left(\ket{2,0}+\ket{0,2}\right),
\eeq
and its wave function reads
\beq
\label{eigenwavedens}
\phi_1(X)\simeq \frac{1}{\sqrt{\pi}}e^{-\frac{X^2}{2}}\left(C_0-\sqrt{2}C_1\left(1-X^2\right)\right).
\eeq
This natural orbit is a superposition of the two first single-particle states of the 2D harmonic oscillator with zero angular momentum, $m=0$, the state $\ket{n=0,m=0}$ and the state $\ket{n=2,m=0}$, thus the wave function has no angular dependence. For the noninteracting case, $C_0=1$ and $C_1=0$, since the particles condense in the ground state of the harmonic oscillator. When the interaction is increased, $C_0$ becomes smaller and $C_1$ starts to increase. In Fig.~\ref{Fig:7}, we plot the wave function of Eq.~(\ref{eigenwavedens}) using the corresponding values of $C_0$ and $C_1$ computed for $N=2,3,4$ and different values of the interaction strength $g$.

%
The advent of correlations beyond mean-field ones should also become 
apparent when computing two-particle correlations. In particular, 
we can evaluate the probability of finding two particles at given 
positions. For simplicity we consider one of them at the origin and 
the second one at a distance $X$. The probability density of finding a 
particle in the space once we have fixed a particle at the center is given 
by $\eta(X)/\rho(0)$ and is normalized to unity (see Appendix~\ref{appendixdenspair}). 
Without interactions, the pair correlation function is proportional 
to the density, since the probability density for finding a particle 
in a particular place is not correlated with the positions of the 
others, see Eq.~(\ref{pairnoint}). In Fig.~\ref{Fig:8}, we show how 
$\eta(X)/\rho(0)$ evolves with increasing the interaction for the 
systems with $N=2$, $3$, and $4$ bosons. In all cases, the central 
peak gets smaller when increasing the interaction, being fairly 
close to zero for $g\simeq 16$. This is in line with the fact that the atoms build correlations to avoid the interaction, e.g. as $g$ is increased the probability of finding two atoms at the same location decreases. In between, next to the center of 
the trap, the function is uniform. When the interaction is strong there is a minimum at the position of the first atom, the probability density $\eta(X)/\rho(0)$ develops a maximum corresponding to the preferred distance between particles. Increasing the number 
of bosons, this maximum shifts towards larger distances. 
\section{Summary and Conclusions}
\label{conclusions}
In this work, we have studied systems of a few number of bosons trapped in an isotropic 2D harmonic trap interacting by a finite-range Gaussian potential.

First, we have explored in detail the noninteracting case, paying particular attention to the degeneracies of the excitation spectrum of the system. In particular, for the $N$-boson case,
we have explained how to compute the degeneracy of the low-energy states which 
is independent of the number of particles.

By means of a direct diagonalization of the Hamiltonian in a truncated space, we have studied the interacting system and we have computed the low-energy spectra for $N=2$, $3$, and $4$ bosons. We have also proposed a variational ansatz with two-body correlations which provides an accurate description of both the energy and the structure of the ground state in the full range of interaction considered.
Center-of-mass and relative excitations are clearly identified in the spectrum. As the interaction is increased, we have shown 
how the ground state and all low lying states tend to saturate as a function of
the interaction strength.

The effect of increasing the interaction on the ground state is twofold. On one side, the density at the center of the trap decreases becoming almost flat in the bulk of the gas, with the cloud thus becoming larger. On the other side, the atoms develop strong two-body correlations to avoid the interaction. This is achieved by building holes in the many-body wave function whenever two atoms are at the same position, as is clearly seen in the computed pair correlations and also on the explicit zeros introduced in our variational wave function. This mechanism is similar to the one present in the Tonks-Girardeau gas in 1D and is also  responsible for the observed saturation of  the energies of the system  as we increase the interaction strength. Finally, the onset of correlations in turn produces fragmentation on the one-body density matrix, which has been shown to increase with the number of particles.

Thus, we have shown that our exact diagonalization method allows one to study interacting bosonic systems in 2D. We are presently implementing this method for spin-orbit coupled bosonic systems.

\vspace*{0.5cm}
\begin{acknowledgments}
The authors thank Th. Busch for his comments on regularization and specially acknowledge B.-G. Englert for sending his notes on this topic. We also would like to show our gratitude to N. L. Harshman for discussions about degeneracy and a careful reading of the manuscript.
The authors acknowledge financial support by grants 2014SGR-401 from Generalitat de Catalunya and FIS2014-54672-P from the MINECO (Spain). P.M. is supported by a FI grant from Generalitat de Catalunya and B.J.-D. is supported by the Ram\'{o}n y Cajal program.
\end{acknowledgments}
\appendix
\section{Computation of degeneracies in the noninteracting limit}
\label{apdeg}
\subsection{The two-boson system}
We compute the degeneracy of each energy level depending on the excitation energy number, $N_E=E-E_0$, for the two-boson system with the possible states labelled using the quantum numbers of Eq.~(\ref{eqboson3}). First, we fix the excitation energy number, $N_E$, and consider it to be even. Then, the values that $n_r$ can take are $n_r=0,2,\,...\, ,N_E$, so $n_r=2k$ with $k=0,1,\,...\,,N_E/2$. Since we have $n_{\rm cm}+n_{r}=N_E$, for each value of $n_r$ there is the corresponding $n_{\rm cm}$. Now, we count the number of states with a given $n_r$ with excitation energy number $N_E$ taking into account the degeneracy due to the quantum numbers $m_{\rm cm}$ and $m_{r}$, that is,
\beq
d_{N_E,k}=(n_{\rm cm}+1)(n_r+1)=(N_E-2k+1)(2k+1).
\eeq
Therefore, we have to sum over $k$ to find the degeneracy. The sum goes from $k=0$ to $k=N_E/2$ if $N_E$ is even and to $k=(N_E-1)/2$ if $N_E$ is odd, which can be generalized using the floor function, summing from $k=0$ to $k=\left \lfloor N_E/2 \right \rfloor$. The degeneracy is
\beq
\begin{gathered}
\label{eqdegcalc}
d^b_{N_E}=\sum_{k=0}^{\left \lfloor N_E/2 \right \rfloor}(N_E-2k+1)(2k+1)=
\\
-\frac{1}{3}\left (\left \lfloor \frac{N_E}{2} \right \rfloor +1 \right)
\\
\times \left[4{\left \lfloor \frac{N_E}{2} \right \rfloor}^2+(2-3N_E)\left \lfloor \frac{N_E}{2} \right \rfloor -3 (N_E+1)\right].
\end{gathered}
\eeq
The previous equation, Eq.~(\ref{eqdegcalc}), for $N_E$ even is
\beq
d^b_{N_E}=\frac{1}{12}\left(N_E+2\right)\left(N_E(N_E+4)+6\right),
\eeq
and for $N_E$ odd is
\beq
d^b_{N_E}=\frac{1}{12}\left(N_E+1\right)\left(N_E(N_E+5)+6\right).
\eeq
For the spatial fermionic states, which are the ones with $m_r=$ odd and antisymmetric  upon exchanging particles $1$ and $2$, we compute the degeneracy analogously, using that $n_r=$ odd,
\beq
\begin{gathered}
\label{eqdegcalcfermions}
d^f_{N_E}=\sum_{k=0}^{\left \lfloor N_E/2 \right \rfloor}(N_E-2k)(2k+2)=
\\
-\frac{1}{3}\left (\left \lfloor \frac{N_E}{2} \right \rfloor +1 \right)
\\
\times \left[4{\left \lfloor \frac{N_E}{2} \right \rfloor}^2+(8-3N_E)\left \lfloor \frac{N_E}{2} \right \rfloor -6 N_E\right].
\end{gathered}
\eeq
\subsubsection{Unperturbed energy states}
We are also interested in knowing the number of states in each energy level with $m_r\neq0$. We compute this number of states subtracting from the total number of degenerate states, $d^b_{N_E}$, the ones with $m_{r}=0$, that is,
\beq
\label{countNCstates}
\begin{gathered}
d^U_{N_E}=d^b_{N_E}-\sum_{k=0}^{\left \lfloor N_E/2 \right \rfloor}(N_E-2k+1)
\\
=\left(-\frac{4}{3}\left \lfloor \frac{N_E}{2} \right \rfloor+N_E+\frac{1}{3}\right)\left \lfloor \frac{N_E}{2} \right \rfloor \left(\left \lfloor \frac{N_E}{2} \right \rfloor+1 \right),
\end{gathered}
\eeq
where we have used Eq.~(\ref{eqdegcalc}). As before, we can separate the case with $N_E$ even,
\beq
d^U_{N_E}=\frac{1}{12}(N_E+2)(N_E+1)N_E,
\eeq
and the case with $N_E$ odd,
\beq
d^U_{N_E}=\frac{1}{12}(N_E+3)(N_E+1)(N_E-1).
\eeq
\section{Computation of the density profile, the pair correlation function and the condensed fraction}
\label{appendixdenspair}
\subsection{The density profile}
\subsubsection{First-quantized density operator}
For a system of $N$ particles, the density operator in first quantization, normalized to unity, is defined as
\beq
\label{defdensityN}
\hat{\rho}(\vec{x})\equiv \frac{1}{N}\sum_{i=1}^N \delta({\vec{x}-\vec{x}_i}).
\eeq
Therefore, the density profile for a given state of a system of $N$ identical bosons, $\Psi(\vec{x}_1,\,...\,,\vec{x}_N)$, would be
\beq
\label{densN}
\begin{gathered}
\rho(\vec{x})=\frac{1}{N}\sum_{i=1}^N \int d\vec{x}_1 \, ... \, d\vec{x}_N \, \delta({\vec{x}-\vec{x}_i}) \left | \Psi(\vec{x}_1,...\, ,\vec{x}_N) \right|^2\\
=\int d\vec{x}_2\, ... \, d\vec{x}_N \,\left | \Psi(\vec{x},\vec{x}_2...,\vec{x}_N) \right|^2 .
\end{gathered}
\eeq
In particular, for a two-boson system in 2D, the previous equation reduces to
\beq
\rho(x,y)=\int_{-\infty}^\infty dx_2\int_{-\infty}^\infty dy_2\left|\Psi \left (x,y,x_2,y_2\right) \right|^2 .
\eeq
We compute the density profile for the general interacting case, in the harmonic trap, for the ground state of the system as
\beq
\label{dens1}
\rho(x,y)=\int_{-\infty}^\infty dx_2\int_{-\infty}^\infty dy_2 \frac{2}{\pi}e^{-\frac{1}{2}\left(\vec{x}+\vec{x}_2\right)^2}\left|f \left (|\vec{x}-\vec{x}_2|\right) \right|^2 ,
\eeq
 where we have made use of the explicit form of the many-body wave function of the ground state,
 \beq
 \label{groundx1x2}
\Psi \left (\vec{x}_1,\vec{x}_2\right)=\sqrt{\frac{2}{\pi}}e^{-\frac{1}{4}\left(\vec{x}_1+\vec{x}_2\right)^2}f\left(\left|\vec{x}_1-\vec{x}_2\right|\right).
\eeq
This way of writing the wave function of the ground state is equivalent to separate the center-of-mass part from the relative part.
Using the change of variables $\vec{r}=\vec{x}-\vec{x}_2$ and polar coordinates in Eq.~(\ref{dens1}), we express the density as
\beq
\label{densityyy}
\begin{gathered}
\rho(x,y)=
\frac{2}{\pi}e^{-2(x^2+y^2)} \int_{0}^\infty r\,  dr \, e^{-\frac{r ^2}{2}} \left|f \left (r \right) \right|^2 \\ \times \int_{0}^{2\pi} d\varphi \, e^{-2r \left(x\cos \varphi+y\sin \varphi\right)}=4e^{-2(x^2+y^2)} \\
\times \int_0^{\infty}r \,dr  \, e^{-\frac{r ^2}{2}}\left|f \left (r \right) \right|^2 \mathcal{I}_0\left(2r \sqrt{x^2+y^2}\right).
\end{gathered}
\eeq
We have used that
\beq
\int_0^{2\pi} d\varphi \, e^{A\cos \varphi+B\sin \varphi}=2\pi\, \mathcal{I}_0\left(\sqrt{A^2+B^2}\right),
\eeq
where $\mathcal{I}_0$ is a modified Bessel function.
Notice that, as we would expect, in Eq.~(\ref{densityyy}) we have demonstrated that the density only depends on the radial coordinate $ X\equiv \sqrt{x^2+y^2}$, and we can rewrite that equation as
\beq
\rho(X)=4e^{-2X^2}\int_0^{\infty}r  dr  \, e^{-\frac{r ^2}{2}}\left|f \left (r \right) \right|^2 \mathcal{I}_0\left(2rX\right).
\eeq
This result is valid not only for our Gaussian-shaped potential but also for any potential dependent only on the modulus of the relative coordinate. In these other cases, the explicit form of the interaction defines the relative wave function $f(r)$.
In the noninteracting case, we can compute the integral analytically, by substituting the explicit form of $f_0(r)$,
\beq
\label{relpartf0}
f_0(r)=\frac{1}{\sqrt{2\pi}}e^{-\frac{r^2}{4}},
\eeq
and we recover the known result,
\beq
\label{dens0}
\begin{gathered}
\rho_0(X)=\frac{2}{\pi}e^{-2X^2}\int_0^{\infty}r  dr  \, e^{-r^2} \mathcal{I}_0\left(2rX\right)\\
=\frac{1}{\pi}e^{-X^2}=\left|\varphi_0(X)\right|^2,
\end{gathered}
\eeq
where $\varphi_0(X)$ is the wave function of the single-particle ground state of the 2D harmonic oscillator. The previous result, $\rho_0(X)=\left|\varphi_0(X)\right|^2$, is also valid for the case of $N$ noninteracting bosons in the 2D harmonic potential, since the many-body wave function factorizes, $\Psi_0(\vec{x}_1,\vec{x}_2...,\vec{x}_N)=\varphi_0(\vec{x}_1)\,...\,\varphi_0(\vec{x}_N)$. We recover the previous result replacing the factorized wave function into Eq.~(\ref{densN}).
\subsubsection{Second-quantized density operator}
For our numerical computations, we make use of the second-quantized form of the density operator,
\beq
\label{2ndqdens}
\hat{\rho}=\frac{1}{N}\sum_{i,j=1}^M\hat{a}^\dagger_i\hat{a}_j\psi^*_i(\vec{x})\psi_j(\vec{x}).
\eeq
For a state written in our Fock basis, Eq.~(\ref{Fockbasis}), as
\beq
\label{stateFock2}
\ket{\Psi}=\sum_{k=1}^{D^M_N} \alpha_k \ket{k},
\eeq
where the index $k$ labels each state of the basis, $\ket{k}=\ket{n_1,\,...\,,n_M}$, the density profile is computed as
\beq
\label{dens2ndstate}
\rho(\vec{x})=\frac{1}{N}\sum_{k',k=1}^{D^M_N}\sum_{i,j=1}^M \psi^*_i(\vec{x})\psi_j(\vec{x}) \alpha^*_{k'}\alpha_{k}\bra{k'}\hat{a}^\dagger_i\hat{a}_j\ket{k},
\eeq
where $\psi_i(\vec{x})$ are the single-particle eigenstates of the 2D harmonic oscillator.
\subsection{The pair correlation function}
The pair correlation operator, normalized to unity, for a system of $N$ particles reads
\beq
\hat{\eta}(\vec{x},\vec{x}')\equiv \frac{1}{N(N-1)}\sum_{i=1}^N\sum_{j\neq i}^N \delta (\vec{x}-\vec{x}_i)\delta(\vec{x}'-\vec{x}_j),
\label{pairN}
\eeq
from which we obtain the pair correlation function for a state of the $N$-boson system, $\Psi(\vec{x}_1,\,...\,,\vec{x}_N)$, as
\beq
\begin{gathered}
\label{pairNfunction}
\eta(\vec{x},\vec{x}')=\frac{1}{N(N-1)}\sum_{i=1}^N\sum_{j\neq i}^N \int d\vec{x}_1 \, ... \, d\vec{x}_N \\
\times \delta (\vec{x}-\vec{x}_i)\delta(\vec{x}'-\vec{x}_j)\left|\Psi(\vec{x}_1,\,...\,,\vec{x}_N)\right|^2\\
=\int d\vec{x}_3 \, ... \, d\vec{x}_N \left|\Psi(\vec{x},\vec{x}',\vec{x}_3\,...\,,\vec{x}_N)\right|^2.
\end{gathered}
\eeq
For the particular case of the ground state of two bosons in 2D, we have
\beq
\label{paircorrel1}
\eta \left (\vec{x},\vec{x}'\right) = \left|\Psi \left (\vec{x},\vec{x}'\right) \right|^2 ,
\eeq
where $\Psi \left (\vec{x},\vec{x}'\right)$ is the corresponding wave function, Eq.~(\ref{groundx1x2}).
For the noninteracting case, in the harmonic trap, we know the function of the relative part, Eq.~(\ref{relpartf0}). In that case, the pair correlation function is
\beq
\eta_{0}\left (\vec{x},\vec{x}'\right) =\frac{1}{\pi^2}e^{-\vec{x}^2}e^{-\vec{x}'^2}=\left|\varphi_0(\vec{x})\right|^2 \left|\varphi_0(\vec{x}')\right|^2.
\eeq
The last result is also valid for the system of $N$ bosons, because then we can factorize, $\Psi_0(\vec{x}_1,\vec{x}_2...,\vec{x}_N)=\varphi_0(\vec{x}_1)\,...\,\varphi_0(\vec{x}_N)$, and replace the wave function into Eq.~(\ref{pairNfunction}) in order to find the same result.

Now, we fix one particle at the origin, and compute the function
\beq
\eta\left(x,y\right)\equiv\eta (\vec{x},\vec{0})=\frac{2}{\pi}e^{-\frac{1}{2}\left(x^2+y^2\right)}\left|f(\sqrt{x^2+y^2})\right|^2.
\eeq
Notice that the previous function depends only on the radial coordinate $X\equiv \sqrt{x^2+y^2}$, so we can write
\beq
\eta (X)=\frac{2}{\pi}e^{-\frac{1}{2}X^2}\left|f(X)\right|^2.
\eeq
Again, for the noninteracting case we have an analytical expression for the previous function, that reads
\beq
\label{pairnoint}
\eta_0(X)=\frac{1}{\pi^2}e^{-X^2},
\eeq
and is proportional to the density, Eq.~(\ref{dens0}).

The probability density of finding a particle in the space once we have found a particle at the origin is given by the quantity $\eta(X)/\rho(0)$. We verify its normalization to unity in the general case,
\beq
\label{probdenspair}
\int d\vec{x}\,\frac{\eta(\vec{x},\vec{0})}{\rho(\vec{0})}=\frac{\int d\vec{x} \,d\vec{x}_3 \, ... \, d\vec{x}_N  \left|\Psi(\vec{x},\vec{0},\vec{x}_3\,...\,,\vec{x}_N)\right|^2}{\int d\vec{x}_2 \, ... \, d\vec{x}_N \left|\Psi(\vec{0},\vec{x}_2,\vec{x}_3\,...\,,\vec{x}_N)\right|^2}=1,
\eeq
where we have used that all the particles are identical, Eq.~(\ref{densN}) and Eq.~(\ref{pairNfunction}).
\subsubsection{Second-quantized pair correlation operator}
The second-quantized form of the pair correlation operator is
\beq
\label{2ndqpair}
\hat{\eta}=\frac{1}{N(N-1)}\sum_{i,j,p,q=1}^M\hat{a}^\dagger_i\hat{a}^\dagger_p\hat{a}_j\hat{a}_q\psi^*_i(\vec{x})\psi^*_p(\vec{x}')\psi_j(\vec{x})\psi_q(\vec{x}').
\eeq
For a state written in our Fock basis, Eq.~(\ref{Fockbasis}), as
\beq
\label{stateFock}
\ket{\Psi}=\sum_{k=1}^{D^M_N} \alpha_k \ket{k},
\eeq
where the index $k$ labels each state of the basis, $\ket{k}=\ket{n_1,\,...\,,n_M}$, the pair correlation function is computed as
\beq
\label{pair2ndstate}
\begin{gathered}
\eta(\vec{x},\vec{x}')=\frac{1}{N(N-1)}\sum_{i,j,p,q=1}^M \psi^*_i(\vec{x})\psi^*_p(\vec{x}')\psi_j(\vec{x})\psi_q(\vec{x}')
\\
\times \sum_{k',k=1}^{D^M_N}\alpha^*_{k'}\alpha_{k}\bra{k'}\hat{a}^\dagger_i\hat{a}^\dagger_p\hat{a}_j\hat{a}_q\ket{k},
\end{gathered}
\eeq
where $\psi_i(\vec{x})$ are the single-particle eigenstates of the 2D harmonic oscillator.
\subsection{The condensed fraction}
The degree of condensation is characterized using the one-body density matrix,
\beq
\rho_{i,j}^{\ket{\Psi}}\equiv \frac{1}{N}\bra{\Psi}\hat{a}^{\dagger}_i \hat{a}_j \ket{\Psi},
\label{eqonebody}
\eeq
where, $i,j=1,\,...\,,M$. Diagonalizing this matrix, its eigenvalues $n_i$ are computed, which are the occupations of the corresponding singe-particle eigenstates $\ket{\phi_i}$. The state $\ket{\Psi}$ is fully condensed when $\ket{\Psi}=\ket{\phi_1}^{\otimes N}$ and then, the one-body density matrix has only a single nonzero eigenvalue, $n_1=1$. If there is fragmentation in the system, the highest eigenvalue $n_1<1$, due to the normalization, $\sum_{i=1}^M n_i=1$.
\section{Computation of the integrals of interaction for the second-quantized Hamiltonian}
\label{apintegrals}
We make an effort to find an analytic expression for the integrals of the interaction part because, in this way, we avoid computing a lot of 4-dimensional integrals numerically, which would mean needing more computational time in order to achieve a good precision before any other calculation. With our method, we have a fast and accurate subroutine that computes $V_{i,j,k,l}$.

In order to compute the integrals, we write explicitly the single-particle wave functions corresponding to the $i^{th}$ eigenstate of the single-particle Hamiltonian,
\beq
\label{eqB1}
\psi_{i(n_x,n_y)}(x,y)=N_{n_x}N_{n_y}H_{n_x}(x)H_{n_y}(y)e^{-\frac{x^2+y^2}{2}},
\eeq
with $H_n(x)$ the Hermite polynomials and the normalization constant
\beq
\label{eqB2}
N_{n}=\left(\frac{1}{\sqrt{\pi}2^{n} n !}\right)^{1/2}.
\eeq
The Hermite polynomials are written in series representation as
\beq
\label{eqgaussMB3}
H_n(x)=\sum_{m=0}^{\left \lfloor n/2 \right \rfloor} \frac{n! (-1)^m 2^{n-2m}}{m! (n-2m)!}x^{n-2m},
\eeq
where $\left \lfloor n/2 \right \rfloor$ indicates the floor function of $n/2$.
We replace Eq.~(\ref{eqB1}) into Eq.~(\ref{eqMB4}) in order to obtain
\beq
\label{eqgaussMB6}
V_{i,j,k,l}=\frac{1}{\pi s^2}\prod_{i=1}^4 N_{n_{xi}}N_{n_{yi}}I_{xx'}I_{yy'},
\eeq
with
\beq
\begin{split}
\label{eqgaussMB8}
I_{xx'}&=\int_{-\infty}^{\infty} dx' H_{n_{x2}}(x')H_{n_{x3}}(x')e^{-A{x'}^2}
\\
&\times \int_{-\infty}^{\infty} dx H_{n_{x1}}(x)H_{n_{x4}}(x)e^{-Ax^2+Bx}
\\
&=\int_{-\infty}^{\infty} dx' H_{n_{x2}}(x')H_{n_{x3}}(x') e^{-A{x'}^2} I_{x}(x'),
\end{split}
\eeq
with the definitions
\beq
\label{eqgaussMB91}
A\equiv1+\frac{1}{s^2},
\eeq
\beq
\label{eqgaussMB92}
B\equiv \frac{2x'}{s^2},
\eeq
and analogously for $I_{yy'}$. Now, we use the series representation of the Hermite polynomials, Eq.~(\ref{eqgaussMB3}), to compute the integral $I_x(x')$,
\beq
\begin{gathered}
\label{eqgaussMB10}
I_x(x')=\sum_{k_1=0}^{\left \lfloor n_{x1}/2 \right \rfloor}\sum_{k_4=0}^{\left \lfloor n_{x4}/2 \right \rfloor} \frac{n_{x1}!n_{x4}! (-1)^{k_1+k_4} 2^{Q}}{k_1!k_4! (n_{x1}-2k_1)!(n_{x4}-2k_4)!}
\\
\times \int_{-\infty}^{\infty}  x^{Q}e^{-Ax^2+Bx} dx
\\
=\sum_{k_1=0}^{\left \lfloor n_{x1}/2 \right \rfloor}\sum_{k_4=0}^{\left \lfloor n_{x4}/2 \right \rfloor} \frac{n_{x1}!n_{x4}! (-1)^{k_1+k_4} 2^{Q}}{k_1!k_4! (n_{x1}-2k_1)!(n_{x4}-2k_4)!}
\\
\times\, i^{-Q}A^{-\frac{Q+1}{2}}\sqrt{\pi} e^{\frac{B^2}{4A}} U \left (-\frac{Q}{2};\frac{1}{2};\frac{-B^2}{4A} \right)
\\
=\sqrt{\frac{\pi}{A}}e^{\frac{B^2}{4A}}\sum_{k_1=0}^{\left \lfloor n_{x1}/2 \right \rfloor}\sum_{k_4=0}^{\left \lfloor n_{x4}/2 \right \rfloor}\sum_{m=0}^{\left \lfloor Q/2 \right \rfloor} \frac{n_{x1}!n_{x4}! }{k_1!k_4! (n_{x1}-2k_1)!}
\\
\times \frac{(-1)^{k_1+k_4}Q!}{ (n_{x4}-2k_4)!m!(Q-2m)!A^{Q-m}}B^{Q-2m},
\end{gathered}
\eeq
where $U \left (-\frac{Q}{2};\frac{1}{2};\frac{-B^2}{4A} \right)$ is a confluent hypergeometric function of the second kind that we have expressed in series and $ Q \in \mathbb{N} $ is defined as
\beq
\label{eqgaussMB11}
Q\equiv n_{x1}+n_{x4}-2k_1-2k_4.
\eeq
The next step is computing the integral in Eq.~(\ref{eqgaussMB8}) by replacing the explicit form of $I_x(x')$, Eq.~(\ref{eqgaussMB10}). First, we notice that depending on the parity of the integrand, the integral will be zero since we integrate in a symmetric interval. The possible situations are
\beq
\label{eqgaussMB12}
 \begin{cases} 
      I_{xx'}=0 & n_{x1}+n_{x2}+n_{x3}+n_{x4} \,\, odd \\
      I_{xx'}\ne 0 & n_{x1}+n_{x2}+n_{x3}+n_{x4} \,\, even.
   \end{cases}
\eeq
In the second case, we compute the integral replacing again the Hermite polynomials by their series representation and substituting (\ref{eqgaussMB10}) into (\ref{eqgaussMB8}),
\beq
\begin{gathered}
\label{eqgaussMB13}
I_{xx'}=\int_{-\infty}^{\infty} dx' H_{n_{x2}}(x')H_{n_{x3}}(x') e^{-A{x'}^2} I_{x}(x')
\\
=\sum_{k_1=0}^{\left \lfloor n_{x1}/2 \right \rfloor}\sum_{k_2=0}^{\left \lfloor n_{x2}/2 \right \rfloor}\sum_{k_3=0}^{\left \lfloor n_{x3}/2 \right \rfloor}\sum_{k_4=0}^{\left \lfloor n_{x4}/2 \right \rfloor}\sum_{m=0}^{\left \lfloor Q/2 \right \rfloor} \prod_{i=1}^4\frac{n_{xi}!}{k_i!(n_{xi}-2k_i)!}
\\
\times \sqrt{\frac{\pi}{A}} \frac{ Q!(-1)^{\sum_{j=1}^4 k_j}2^{Q'}}{m!(Q-2m)!A^{Q-m}s^{2Q-4m}}\int_{-\infty}^{\infty} {x'}^{Q'}e^{-A'{x'}^2} dx'
\\
=\sum_{k_1=0}^{\left \lfloor n_{x1}/2 \right \rfloor} \sum_{k_2=0}^{\left \lfloor n_{x2}/2 \right \rfloor}\sum_{k_3=0}^{\left \lfloor n_{x3}/2 \right \rfloor}\sum_{k_4=0}^{\left \lfloor n_{x4}/2 \right \rfloor} \sum_{m=0}^{\left \lfloor Q/2 \right \rfloor}\prod_{i=1}^4\frac{n_{xi}!}{k_i!(n_{xi}-2k_i)!}
\\
\times \sqrt{\frac{\pi}{A}}\frac{ Q!(-1)^{\sum_{j=1}^4 k_j}2^{Q'}A'^{-\frac{Q'+1}{2}}\Gamma\left ( \frac{Q'+1}{2}\right)}{m!(Q-2m)!A^{Q-m}s^{2Q-4m}},
\end{gathered}
\eeq
with the definitions
\beq
\label{eqgaussMB141}
A'\equiv A-\frac{1}{As^4},
\eeq
\beq
\label{eqgaussMB142}
Q'\equiv \sum_{i=1}^4 \left (n_{xi}-2k_i\right)-2m.
\eeq

The expression is analogous for $I_{yy'}$ and all the sums that appear are finite and have few terms when $n_{xi}$ are small. Now, knowing the form of $I_{xx'}$ and $I_{yy'}$ we have $V_{i,j,k,l}$. Moreover, many of the integrals are zero
\beq
\label{eqgaussMB15}
 \begin{cases} 
      V_{i,j,k,l}=0 & \sum_{i=1}^4 n_{xi} \, \, odd\,\,\, or \,\,\,  \sum_{i=1}^4 n_{yi} \, \, odd\\
      V_{i,j,k,l}\ne 0 & \sum_{i=1}^4 n_{xi} \, \, even\,\,\, and \,\,\,  \sum_{i=1}^4 n_{yi} \, \, even,
   \end{cases}
\eeq
and we also take profit from the symmetries of $I_{xx}(n_{x1},n_{x2},n_{x3},n_{x4})$, which verifies
\beq
\begin{gathered}
I_{xx}(n_{x1},n_{x2},n_{x3},n_{x4})=I_{xx}(n_{x4},n_{x2},n_{x3},n_{x1})\\
=I_{xx}(n_{x1},n_{x3},n_{x2},n_{x4})=I_{xx}(n_{x4},n_{x3},n_{x2},n_{x1}).
\end{gathered}
\eeq
Therefore, we are computing four integrals at the same time.

\end{document}